\documentclass[12pt,preprint]{aastex}

\slugcomment{ }

\lefthead{Pentericci et al.}

\righthead{Near infrared properties of HzRGs.}

\begin{document}

\title{NICMOS observations of high redshift radio galaxies: witnessing the formation of bright elliptical galaxies?}

\author{L. Pentericci\altaffilmark{1,3},P.J.McCarthy\altaffilmark{2},H.J.A. R\"ottgering\altaffilmark{3},
G.K.Miley\altaffilmark{3}, W.J.M.van Breugel\altaffilmark{4} and
R. Fosbury\altaffilmark{5}} 
\altaffiltext{1}{Max Planck Institute f\"ur Astronomie, Konigstuhl 17, D-69117 Heidelberg, Germany}\altaffiltext{2}{The Observatories of the Carnegie Institution of
Washington,813 Santa Barbara Street, Pasadena, CA 91101 USA}\altaffiltext{3}{Leiden
Observatory, P.O.~Box 9513, 2300 RA Leiden, 
The Netherlands}\altaffiltext{4}{Institute of Geophysics and 
Planetary Physics, Lawrence Livermore National laboratory,PO Box 808,
Livermore, CA 94459 USA}\altaffiltext{5}{Space Telescope European Coordinating Facility, 
Karl-Schwarschild-Strasse 2, 85748 Garching, Germany}
















\begin{abstract}
We present the results of a near infrared imaging program  of a
sample of 19 radio galaxies with redshift between 1.7 and 3.2,
using the NICMOS Camera 2 on the Hubble Space Telescope. The
galaxies were observed in H-band which, for 18  of the 19 targets,
samples the rest-frame optical emission longwards of the 4000 \AA \space
break. For many sources this band contains emission lines,
but we estimated that this causes relatively little confusion in most cases.
\\
The high angular resolution of the HST allows a detailed study of
the host galaxies. The images show a wide range of morphologies,
including (i) compact systems, (ii)
systems with substructures such as multiple emission peaks  and
(iii) systems comprised of several components spread over areas of up
to 100 kpc. Three galaxies appear unresolved and in two others a
nuclear point source dominates the emission in the central region.
\\
The morphologies of some of the lowest redshift targets
are  well represented by de Vaucouleurs
profiles, consistent with them being elliptical
galaxies. Their average effective radius derived is a factor of 2 smaller
than that of z $\sim$ 1 3CR radio galaxies at similar restframe wavelength.
\\
The near infrared continuum light is generally well  aligned with
the radio axis and the aligned light is very red, with typical V-H
colors of $\sim$ 3.5--4. For several galaxies where WFPC2 V or R-band
images were available we computed a high resolution map of
the optical-to-infrared spectral indices: all multi-component
systems present net color differences between the
various clumps, and we argue that most probably the
continuum emission has a stellar origin. Indicative ages of these stellar 
populations, as determined 
by the amplitude of the 4000 \AA \space break vary between 0.5 and 1.3 Gyrs.
\\
Finally in many of the small NICMOS frames we observe 
nearby faint objects close to the high redshift radio galaxies. The
number density of these faint objects is slightly higher
than that observed in the deep NICMOS parallel observations of
random fields: furthermore these  objects tend to be
aligned with the direction of the main axis of the radio sources,
suggesting that they may be related to the presence of the AGN.
\end{abstract}





\section{Introduction}
The study of the early  Universe has received a considerable
boost in the past few years, due to the development of new techniques
for finding large numbers of galaxies at high redshift
(e.g. U and B band dropouts, Steidel et al.\ 1996).
Although radio galaxies are no longer the only class
of well studied high redshift galaxies,
they remain of exceptional cosmological
interest, since they are likely to be amongst the most massive galaxies 
known in the early Universe  and may be  the progenitors of brightest
cluster galaxies observed at low redshift.
\\
Their large masses can be inferred from
their K--band luminosities and the near-IR K-z Hubble diagram
(e.g. van Breugel et al.\ 1998).
Furthermore there is evidence that they are undergoing vigorous star formation:
at least in some high redshift radio galaxies (HzRGs) the UV continuum is 
dominated 
by young stars,
with estimated star-formation
(SF) rates of up to 1000 M$_\odot$ yr$^{-1}$ (Dey et al.\ 1997). 
Sub-millimeter continuum dust emission has
been detected in several of the highest
redshift galaxies (e.g. Papadopoulos et al.\ 2000),
also implying  similar large SF rates. These stars are expected
to settle on dynamical time-scales and evolve into fully developed 
elliptical galaxies.
\\
Evidence that HzRGs might be located in proto- cluster environments
includes:
(i) the recent discovery of a megaparsec-scale structure of more than 
15 Ly$\alpha$ emitting galaxies around the radio galaxy MRC 1138$-$262 
at z$=2.156$ (Pentericci et al.\ 2000a); around this same radio galaxy,
the detection of luminous X-ray emission
 which is probably extended and has been attributed to
a hot cluster atmosphere (Carilli et al.\ 1998); 
(ii) large Faraday rotation of the polarized radio emission
indicating that some HzRGs are surrounded by dense hot magnetized cluster gas
(Carilli et al.\ 1997; Pentericci et al.\ 2000b);
(iii) excess of companion galaxies detected along the axes or in the vicinity
of the radio sources (R\"ottgering et al.\ 1996);
(iv) possible excess
of Lyman break selected galaxies in the fields
of several other powerful radio sources
(e.g. Lacy \& Rawlings 1996) and in particular several
candidate companion galaxies
(with two objects spectroscopically confirmed)
in the vicinity of MRC 0316$-$257, at z$=3.14$ (Le Fevre et al.\ 1996;
McCarthy et al.\ 1992), and a number of faint red companions 
of 4C 41.17 at z$=3.8$ detected from deep Keck imaging at 2 $\mu$m 
(Graham et al.\ 1994). 
\\
Having both large stellar masses (e.g. 4C41.17 with
M$ \sim 10^{11} M_{\odot}$, van Breugel et al.\ 1998) and high
star formation rates (see above),
and being located in the densest regions of the early Universe,
it is natural to propose that HzRGs will evolve into present-day 
brightest clusters
galaxies (e.g. Best et al.\ 1998).
\\
\\In previous papers we have described our studies of HzRGs 
with the Hubble
Space Telescope at optical (UV rest-frame) wavelengths
(Pentericci et al.\ 1998 and 1999).
It was found that several HzRGs, such as  MRC 1138$-$262
at z$=2.2$  and 4C41.17 at z$=3.8$ (Dey et al.\ 1997) 
are comprised of  numerous continuum clumps embedded in
giant ($\sim$ 100 kpc) diffuse Ly$\alpha$ halos.
Such morphologies
are strikingly similar to simulations of forming bright cluster
galaxies, made on the basis of the hierarchical models, which
predict that the most massive
systems are assembled from smaller building blocks (e.g. Baron
and White 1987; Dubinski 1998). Indeed the sizes,
profiles and luminosities of the {\it individual} clumps are similar 
to those of the Lyman
break galaxies (LBG) (e.g. Steidel et al.\ 1996, Giavalisco
et al.\ 1996), indicating that HzRGs could be formed by an assembly
of LBGs (Pentericci et al.\ 1998).
\\
The well known alignment observed between the rest-frame UV morphology and 
the radio axis 
(Chambers et al.\ 1987, McCarthy et al.\ 1987) implies that the UV emission is 
strongly effected by the active 
galactic nucleus, e.g. by scattered light from a QSO or a young population of
stars whose formation is triggered by the passage of the radio jet.\ Studies
of the near-infrared morphology (rest-frame optical) are particularly
important to reveal the nature of the older 
stellar population (e.g. Lilly 1988),  although observations of the continuum 
emission are made difficult by the presence of 
strong emission lines in the near-IR bands (Eales \& Rawlings 1993). 
\\
There have been extensive ground--based near-IR
studies of powerful radio galaxies
at intermediate and high redshift (e.g. Eisenhardt and Chokshi 1990; McCarthy et al.\ 1992; Eales et
al.\ 1997; Best et al.\ 1998; van Breugel et al.\ 1998).
The most important results of these searches are: (i)
the morphologies of z $\sim$ 1 galaxies in the near-IR emission are considerably
more relaxed and symmetric than in the optical emission; surface
photometry of 3CR  z $\sim$ 1 galaxies has shown that they
follow a de Vaucouleurs law (Best et al.\ 1998),
implying that they are  dynamically evolved systems (e.g. Rigler and Lilly 
1994)
with inferred stellar  masses of up to  10$^{12} M_{\odot}$;
(ii) 3CR radio galaxies show a
possible excess of emission at large radii, resembling the  halo
that surrounds nearby cD galaxies (Best et al.\ 1998);
(iii) models of spectral energy distributions of radio galaxies
having redshifts between 1 and 3 have been used
to infer the ages of their stellar populations, suggesting ages
in excess of 1 Gyr and  formation redshift of z$\sim$5 or higher
(e.g. McCarthy 1993b, Lilly 1988); (iv)
deep near infrared (K-band) imaging reveal clustering of red galaxies
around some z$ \sim 1$ 3CR galaxies (e.g. Roche et al.\ 1998; Best 2000); (v)
van Breugel et al.\ (1998) found strong evolution of the morphology of HzRGs 
at rest frame visual ($ > 4000$ \AA)
wavelengths, from aligned galaxies at redshift $> 3$ to more
 symmetrical and compact galaxies at z$< 3$.
\\
With NICMOS it is possible to study the near infrared morphology of HzRGs
to a resolution comparable to  that of our optical WFPC2 images, allowing us
to investigate the morphologies of their evolved stellar populations 
at redshift larger than 2.
We can then address questions such as whether there are already well
formed ellipticals at those epochs; the importance 
and the frequencies of merging with
sub-galactic clumps at different epochs; how the presence 
of substructures evolves with redshift;
the relationship between the
radio source and the near infrared emission in the central regions of the 
galaxies. In particular, one can examine to what extent 
the optical/UV alignment effect is also present at
near infrared wavelengths. 
\\
In this paper we present NICMOS observations of a sample of 19 radio galaxies
with 1.8 $\le z \le $ 3.2.
In Section 2 we describe the sample selection and the observations.
Section 3 is devoted to the presentation of the results, and a
 description of the morphologies of the individual objects.
In Section 4 we analyze the morphologies of the galaxies
 and their optical-to-infrared colors; 
we then discuss the implication of our NICMOS results for 
the study of the alignment effect and we investigate 
 the statistics of (aligned) companion
galaxies around the targets.

\section{Observations}
\subsection{Sample selection}
 Two sub-samples of objects were observed. The first sub-sample consists of 14
radio galaxies
from the Molonglo Reference Catalog (MRC) (McCarthy et al.\ 1996).
The MRC/1Jy sample was
defined simply using a flux density limit at 408 MHz  of S$_{408} \ge$ 0.95 Jy,
 within a restricted region of the sky ($-30 <$ DEC$ < -20$, $|b|>15$); 
$99.5$\% of the sources in the catalog have been optically identified, and
46 of the 426 radio galaxies in the sample
have spectroscopically confirmed redshift $1.6 < z < 3.2$
(see McCarthy et al.\ 1990 and
McCarthy et al.\ 1991 for the optical identification and
spectroscopy of the higher redshift galaxies).
The MRC/1Jy survey is one of the largest
complete radio sample for which there is a homogeneous and nearly
complete set of optical and infrared photometry
(see the mentioned papers and McCarthy et al.\ 1992 for infrared observations).
\\
We then added  5 targets selected from a sample of ultra steep spectrum
 radio galaxies (USS).  This sample comprises the largest set of radio
galaxies having redshift larger than 2, which were selected on the basis of
the radio spectral indices ($\alpha \le -1.1$ where $S_\nu=S_0\nu^{\alpha}$).
This selection technique has proven to be the most
effective tool for identifying
such radio galaxies. Several samples of USS sources
(e.g. Chambers et al.\ 1996, R\"ottgering et al. 1994)
have led to the discovery of more than 80 radio galaxies with z$ >$ 2, about
60$\%$ of the 150 such sources known to date (e.g. van Breugel et al. 1999).
All the highest redshift radio galaxies (z $>$ 3.5) have been found using
this technique (e.g.\ van Breugel et al.\ 1999).
The USS sample covers a larger area in the sky than the MRC sample, and
allows  us to probe a wider range of radio sources parameters such as their
luminosity and size. Furthermore selecting sources from the USS sample
will give a larger overlap  with the existing data base of WFPC2 images
(Pentericci et al.\ 1999).
\\
The final sample consists of 19 radio galaxies, uniformly
spanning a redshift range between z $=$ 1.68 and z $=$ 3.13, and
having a large  range of  properties,
e.g. a range of nearly 3 magnitudes in the infrared luminosity
and a factor of 8 in radio power. The radio galaxies are listed in 
Table \ref{tbl-1}: for each object we report the redshift, 
the position of the 
radio core, the total flux at 4.7 GHz,  the spectral index between 4.7 and 
8.2 GHz, $\alpha_{8.2}^{4.7}$, the total size in arcseconds and an 
indication of the radio morphology.

\subsection{Observations and data reduction}
\subsubsection{HST/NICMOS}
In Table \ref{tbl-2} we 
list the radio galaxies observed and various observational 
parameters. 
All objects were observed with NICMOS Camera 2, which has a plate scale
of $0.0762 \times 0.0755$ arcseconds per pixel and a total field of view of 
19.2$'' \times$ 19.2$''$.
The filter used was the F160W filter, which is centered at 1.6 $\mu$m,
has a bandwidth of 0.4$\mu$m and has the minimum
background amongst the available NICMOS filters; we also used 
the F165M filter which is centered at 1.7$\mu$m and has a bandwidth of
0.2$\mu$m.
We had initially selected this second filter for many of the objects, because 
it would have provided a bandpass free from line emission
for most of the galaxies in the sample, in particular those in the lowest
and in the highest redshift ranges (z$ <$ 2.09 and z $>$2.6).
Unfortunately using this filter the  response was limited  by
readout-noise and dark current and we had to change 
our initial observing strategy to use the F160W filter.
The wider  filter bandwidth  included bright emission lines for many 
of the sources (see Table \ref{tbl-2}), but we estimate that in most cases  this causes 
relatively little confusion in determining their real continuum morphologies
(however see next section).
The high sensitivity of this filter allowed the 
galaxies to be mapped within the allocated 2 to
5 orbits per object.
\\
Each orbit included  two 1026 seconds exposures using the MIF1024 Multiaccum
exposure sequence and one shorter exposure using the STEP64 sampling sequence.
The exposure times for the STEP64 samples were 384, 411, 512 or 576 seconds
depending on the orbit.
The total integration time for each galaxy is given in Table \ref{tbl-2}.
To facilitate removal of residual flat field and imperfection of the detector,
the targets were offset by  $\sim 3''$ between each exposure, giving a
grid of 15, 12, 9 or 6 exposures.
\\
The NICMOS imaging data were processed in the following manner: the zero
level determined from the first read of each exposure sequence was subtracted
from all other exposures; then a scaled version of the best dark exposure 
provided by the HST
archive was subtracted. The resulting sequences of ``read''
in each Multiaccum or STEP64 series  were  then used to produce a masked,
linearized and, to first order, cosmic ray
rejected image using an algorithm written by McLeod (1997).
Then for the exposures  which were obtained before August 1997
the pedestal level (i.e. significant fluctuations of the zero
level between sequences)
was subtracted separately from each quadrant of the HgCdTe 
detector
and again the cosmic ray rejection algorithm was used.
For the observations taken after August 1997 the last two steps could be
skipped.
Finally, the images were flat fielded using the best flat field provided
from the HST archive.
The images were then registered using fractional pixel shifts, 
and were combined
using a mask for removal of bad pixels and hot pixels (this mask was 
constructed directly using the observation frames)
and a multi pass $\pm 3 \sigma$ cosmic ray rejection algorithm. In combining
the images, we weighted them using  their exposure times.

\placetable{tbl-1}

\placetable{tbl-2}
\subsubsection{Supporting ground-based observations}
For 3 radio galaxies in the sample (MRC 0156$-$252,
MRC 0406$-$244 and MRC 2104$-$242)
we  gathered  extensive  ground-based optical and near-IR
data during the last few years. In Table \ref{tbl-3} we summarize
the relevant observations. Some results on
MRC 0406$-$244 were already published by Rush et al.\ 1997.
\\
For the spectroscopic observations at the ESO/NTT Telescope, 
the detector was a Tektronix CCD with
$1024^2$ pixels and a scale along the slit of 0.37$''$ per pixels.
The CCD was binned by a factor of two in the wavelength direction.
Using a 2.5$''$ wide slit with ESO grating $\#3$ we achieved a spectral
resolution of 2.8 $\rm \AA$ (full width at half maximum, FWHM).\ The
raw spectra were bias subtracted and flat fielded; the sky
contribution was then removed by subtracting a sky-spectrum obtained 
avoiding the regions where the target was
positioned; finally  wavelength calibration was performed by measuring
the position on the CCD of known lines from a He-Ar calibration lamp,
fitting a polynomial function to these data and applying the resultant
calibration factors. The accuracy of the calibration is
better than 0.3 \AA.
 \\
The J and K images were obtained with the Las Campanas du Pont 2.5m
telescope and its $256 \times 256$ HgCdTe array camera, 
which has a pixel scale of 0.35$''$.\ The integration times
ranged from 4000 to 9000 seconds and the objects were moved on the array
every 150 seconds. The data were calibrated with standard stars from
Elias et al.\ (1983). The reduction followed standard techniques.
The $r$, I and Ly$\alpha$ images were also obtained at Las Campanas using
 a $800 \times 800$ Texas Instruments CCD detector: the  exposures were 
binned 2$\times$2 resulting in a pixel scale of 0.332$''$. The Ly$\alpha$
images were obtained with interference filters having 1\% band-widths.
Several 1800 second exposures of each field were obtained, with
approximately $10^{''}$ offsets between each image.
\\
All data were reduced using standard IRAF procedures.
The various images were then registered with the NICMOS frames using the position
of several point sources in the field, with the AIPS
(Astronomical Image Processing System) tasks XTRAN and HGEOM, assuming a 
linear transformation.
The accuracy of the registration is about 0.2$''$.
\\
All the radio galaxies  with the exceptions listed below were imaged
with the VLA  in A-array as part of a high resolution, multi-frequency
radio polarimetric study carried out on a large sample of HzRGs; for 
further details about the observations see Carilli et al.\ (1997).
For a subset of galaxies, new VLA observations were carried out: 
the radio galaxies MRC 0324$-$228 and MRC 0350$-$279 were observed in the
B-array configuration at 4.85 GHz (C-band). The radio galaxies
MRC 0152$-$209, MRC 1017$-$220, MRC 2048$-$272 and MRC 2224$-$273 were 
observed in the A-array configuration
as part of a new  high resolution multi-frequency radio survey.
Details of these  observations have been  be presented
elsewhere (Pentericci et al.\ 2000b).

\subsection{Emission-line contamination}

As explained earlier we observed most of our targets with
 the F160W broad band filter.\ This resulted in the unavoidable inclusion of
bright  emission lines normally found in radio galaxies.
For the lowest redshift object (MRC 2224$-$273 at z $=$ 1.68), the strong 
emission line H$\alpha$ falls in the
observed wavelength range; for the objects in the redshift range
1.9$ < z < $2.6,
[OIII] and H$\beta$ become important; finally for the higher redshift objects
(z $\ge 2.9$) the emission line [OII] is within the observed wavelength range.
In Table \ref{tbl-2} we have listed, for each object, the lines that can contribute
substantially to the continuum band flux and the total estimated contribution
to the continuum flux.
\\
Since we do not have direct
measurements for the line fluxes and their equivalent widths, 
we have estimated the total
contaminating flux using other measured lines
(in most cases Ly$\alpha$ and/or H$\alpha$).
We have used the emission line ratio reported by McCarthy (1993) and 
Eales \& Rawlings (1996), i.e. Ly$\alpha$/[OII]$=5$, [OIII]/[OII]$=3$, 
Ly$\alpha$/H$\alpha$ $=1.6$. Note however that these are 
only average ratios. For example, Eales \& Rawlings (1993) 
find that the line emission ratios can
vary significantly (within a factor of 10) from object to object, due to 
the different
physical conditions of the gas and/or the presence of dust which
can attenuate the Ly$\alpha$ line (e.g. in USS 0211$-$122, van Ojik et
al. 1994). Another important source of uncertainty is  the
difference in the apertures used to derive the continuum and
line fluxes: the continuum flux was measured through a fixed 4$''$
diameter aperture, while the line fluxes (mostly taken from the
literature) were measured within different apertures and with slits 
of different sizes.
When the apertures used to determine the emission line flux were
known (in about half of the cases) we corrected for this
by simply scaling the line flux to an aperture of 4$''$. Therefore
the values for the emission line contamination given in Table \ref{tbl-2} 
are only indicative. 
Most important, the line contribution may vary spatially and some 
parts of the galaxies  might be more effected than others, 
with a much higher line contribution than that listed in Table \ref{tbl-2}.

\subsection{Astrometry}
The coordinate frame for the NICMOS images determined from the
image header information has uncertainties of the order of $0.6''$
due to the uncertainty in the position of the guide star (HST data handbook).
Since the optical galaxies are generally clumpy on a scale
of a few tenths of arcsecond, it is important to obtain the better relative
registration between the radio and the optical images
for a detailed inter-comparison.
\\
Our radio maps have a typical resolution of 0.2$''$
which is comparable to that of the NICMOS images. The exceptions are the radio
 maps of MRC 0324$-$228
and MRC 0350$-$279 which have a resolution of 1.2$''$.
\\
To align the NICMOS and the radio VLA images  
we overlayed the radio cores 
(for the identification of radio cores see Carilli et al.\ 1997 and Pentericci
et al.\ 2000)
on the peak position of the near 
infrared emission, on  the assumption that
this IR peak coincides with the galaxy nucleus. In these cases 
the estimated uncertainty in the overlay is $\sim 0.1''$. 
In the cases of MRC 0324$-$218, MRC 0943$-$242, USS 1707$+$105 
and MRC 2048$-$272 where  no radio core was detected,
we used the absolute HST and VLA astrometry to align the maps; in these cases 
the accuracy achieved is only $\sim 0.8''$. 

\section{Results}
\subsection{Summary}
In Figures 1 to 19 we present the images of the radio 
galaxies.
In most cases we show grey scale representations of the NICMOS
emission with VLA contours superimposed (unless the radio emission is
unresolved) and a contour map of the continuum emission 
to better delineate the morphology of the central regions
(for all sources except the three unresolved ones). 
\\
In Table \ref{tbl-4} we list the total NICMOS H-band magnitudes 
for each object, derived using a fixed 4$''$ circular aperture.\ The 
errors in the magnitudes, reported in column 5,
are dominated by the inaccuracy in the sky subtraction. 
For those galaxies that have more components, 
we also report the magnitudes of the other clumps brighter than 23.5: 
the letters in Table \ref{tbl-4} refer to the components 
as labeled in the figures. For MRC 1138$-$262 and USS 1707+105 
we kept the same 
nomenclature as previous papers (Pentericci et al. 1998;1999).
\\
The NICMOS H-band $3\sigma$ limiting surface brightnesses achieved ranges from
$\mu_{H(3\sigma)}=24.8$ to $\mu_{H(3\sigma)}=25.1$ (see Table \ref{tbl-2}) 
for the 
objects that were imaged with the F160W
filter and only $\mu_{H(3\sigma)}=23.7$ for those observed with the F165M
filters and for shorter exposure times. In almost all cases the signal to
noise ratio is good and allows us to study in detail the morphology of
the galaxies.  The exceptions are the very faint radio galaxy
MRC 0324$-$228 at z $=$ 1.89 and the highest redshift object
in the sample, MRC 0316$-$257 at z $=$ 3.13.
\\
Our observations show that HzRGs have a wide variety of
near-infrared morphologies.  Of the 19 galaxies, 3 are unresolved
(MRC 0350$-$279, MRC 1017$-$220, USS 2202$+$128); of these, the first 2
are also unresolved in the radio at all frequencies and MRC 1017$-$220
has been classified as a broad line radio galaxy (BLRG) by Kapahi et al.\ (1998)
on the basis of its emission line spectrum. In two other cases, MRC 1138$-$262
and MRC 2025$-$218, a nuclear point source dominates the emission in the central regions, but there is clearly underlying extended emission.\ For
all other systems no nuclear point source contributes
substantially to the emission.
The morphologies of the galaxies vary from compact or unresolved structures
to systems comprised of several components spread over large areas 
(up to 100 kpc).\ A large fraction of the systems 
shows close (within $\sim 5''$) emission
 clumps, that might be part of  the systems, but only in few cases, where 
additional
information is available (e.g.\ narrow band imaging, spectroscopy, etc.)
we can conclude that these small components are physically associated
to the radio galaxies.
\\
Comparison with the radio sources shows that typically the near infrared
emission is well aligned with the radio axis. The alignment effect will be
extensively discussed in Section 4.
\\
In the next subsection we will make some remarks on the individual
objects.
\subsection{Individual objects}
\noindent
{\bf MRC 0140-257  z $=$ 2.64}
\\
The galaxy is shown in Fig.\ 1: it has 2 peaks of emission
with almost the same flux.
The orientation of the line joining these peaks is, within a few degrees, along
the radio axis. It is possible that the observed morphology is due to an obscuring dust lane.
There are a few fainter clumps within 2-3$''$ of the galaxy that could
be physically associated with the system.
\\
\\
{\bf MRC 0152-209  z $=$ 1.89}
\\
The galaxy is shown in Fig.\ 2:
it is elongated and it is surrounded by a halo of fainter emission.
There is a narrow elongated tail that departs from the southern end
of the galaxy. The radio source is unresolved.
\\
\\
{\bf MRC 0156-252   z $=$ 2.09}
\\
This galaxy is shown in Figs.\ 3 and 20a-d
and is discussed more extensively in Section 3.3.
\\
\\
{\bf USS 0211-122  z $=$ 2.34}
\\
The host galaxy (Fig.\ 4) of this very large (134 kpc) radio source
consists of a nucleus
plus faint diffuse emission, having the shape of an arc.
The galaxy is misaligned with respect to the radio axis; however 
the radio source 
shows  a jet feature extending from the core 
towards south, whose curvature suggests that the radio axis might be 
precessing.
\\
\\
{\bf MRC 0316-257   z $=$ 3.13}
\\
The galaxy is shown in Fig.\ 5.
There are two objects separated by only $\sim 2''$.
The identification is the fainter diffuse object to the west.
The  other object could well be at the same redshift. Given
the high redshift of
this radio galaxy, the NICMOS image samples
rest-frame continuum emission that is mostly below the 4000 \AA \space break,
therefore we are not
really imaging the older stellar population.
\\
\\
{\bf MRC 0324-228   z $=$ 1.89}
\\
The galaxy is shown in Fig.\ 6.
There are two faint and diffuse objects separated by only $\sim 1.5''$
which  could be identified as the host galaxy of the radio source.
Ground-based observations do not resolve the 2 components
(McCarthy et al.\ 1996),
which could be part of the same system; alternatively,
one could be the radio galaxy
and the other just a foreground object.
The VLA snapshot radio image at 4.5 GHz
with a resolution of 1.2$''$, shows a simple double radio source with no 
core detected.
The direction of the radio axis is at 25$^{\circ}$ with respect to
the relative  orientation of the two emission clumps.
\\
\\
{\bf MRC 0350-279   z $=$ 1.90}
\\
The identification of this radio source is the unresolved object in 
Fig.\ 7.
There are two faint objects
at a distance of only 6$''$ to the east: one of them
is compact and the other one
diffuse and elongated.
The radio source is unresolved in our VLA B-array snapshot observation,
so we do not show it.
\\
\\
{\bf MRC 0406-244   z $=$ 2.44}
\\
This galaxy is shown in Figs.\ 8 and 21a-d
and is discussed further in Section 3.3.
\\
\\
{\bf MRC 0943-242   z $=$ 2.93 }
\\
This galaxy is shown in Fig.\ 9: it
has a cigar shape and is elongated along the radio axis.
The WFPC2 optical image is remarkably similar. The galaxy is
extensively discussed in McCarthy et al.\ 2000 (in preparation).
\\
\\
{\bf MRC 1017-220  z $=$ 1.77}
\\
The identification of this radio source is the unresolved object in 
Fig. 10.
There are 2 faint diffuse
clumps about 5$''$ away. The radio source is also unresolved.
Kaphai et al.\ (1998) have classified this radio galaxy as a BLRG 
on the basis of its emission line spectrum.
\\
\\
{\bf MRC 1138-262  z $=$ 2.16 }
\\
The central object  of this very large system in Fig.\ 11
has a large contribution from
a nuclear point source, but clearly there is extended
emission below it.
In addition several other components are visible in a region of about
$10'' \times 10''$ around the nucleus: all have
WFPC2 optical counterparts, and are embedded in a giant Ly$\alpha$ halo.
This  galaxy is further
discussed in Pentericci et al.\ 1997 and 1998.
\\
\\
{\bf USS 1410-001   z $=$ 2.33}
\\
This is a very elongated object (Fig.\ 12), consisting of the galaxy core
plus 2 other clear peaks and diffuse emission extending for more
than 4$''$ along the galaxy axis. The  very red (V-H=5.6) and compact
object only 2$''$  from the core, might also be part of the
system.
 The galaxy and the radio source are strongly misaligned:
the angle between the optical and radio axis is nearly 45$^{\circ}$.
However, the northern component of the radio source is curved,  suggesting the
 radio axis might be precessing, in which case the elongated near infrared
emission could be extended along the previous position of the radio jet.
\\
\\
{\bf USS 1707+105   z $=$ 2.35 }
\\
This is a large multiple  system (Fig.\ 13) consisting of 2 or possibly 3
different galaxies (see Pentericci et al.\ 1999)
 separated by 3-4$''$ and located along the radio axis.
The radio core is not detected in the present radio image, however it seems
most likely that the radio emission
comes from the most luminous and larger galaxy, centrally located between the 
two radio lobes
that is indicated with the letter A in Fig.\ 13.
This galaxy is comprised of a main body plus a 
few smaller clumps and shows diffuse emission towards north, along 
the direction of the radio axis, but also perpendicular to it.
\\
\\
{\bf MRC 2025-218   z $=$ 2.63 }
\\
The galaxy is shown in Fig.\ 14.
The emission is dominated by a nuclear point source.
Once this is subtracted there is a large diffuse galaxy 
roughly aligned with
the radio axis.
The radio source is small and the southern radio jet presents 
a sharp bent.
\\
\\
\noindent
{\bf MRC 2048-272   z $=$ 2.06}
\\
The galaxy is shown in Fig.\ 15.
Two bright objects can be seen separated only by 2.5$''$
with a third faint object  $2''$ to the east.
With the present astrometry the correct identification is
probably the central  object. The radio core is undetected 
in our VLA radio images.  
\\
\\
{\bf MRC 2104-242  z $=$ 2.49}
\\
This galaxy is shown in Figs.\ 16 and 23a-c,
and is discussed more extensively in Section 3.3.
\\
\\
{\bf USS 2202+128   z $=$ 2.70}
\\
The object (Fig.\ 17) is not resolved on the present image.
The radio source is small
and s-shaped  suggesting the radio axis might be precessing.
\\
\\
\noindent
{\bf MRC 2224-273   z $=$ 1.68}
\\This galaxy (Fig.\ 18) has a simple pear-shaped morphology. There is
a small compact object only 2.6$''$ away, located
along in the direction of the galaxy main axis.
The radio source is unresolved at all frequencies in our VLA observations
(Pentericci et al.\ 2000b).
\\
\\
{\bf USS 2349+280   z $=$ 2.89}
\\
The galaxy is shown in Fig.\ 19: it consists
of 2 components embedded in a halo of
diffuse emission. The sharpness of the separation between the two components
suggest that is could be due to a dust lane (the NICMOS image samples the
emission interval
 3600-4630 \AA, which can be still effected strongly by dust).
The WFPC optical image is very similar (Chambers et al.\ 1996).

\subsection{Detailed multi-wavelength comparisons}
\noindent
{\bf MRC 0156-252}
\\
In Fig. 20a-c we show the NICMOS image in grey scale overlayed 
with contours from
(a) the I-band emission, (b) the K-band emission and (c) the Ly$\alpha$
narrow band emission.
In Fig. 20d we show in grey scale the radio jets at 8.2 GHz
superimposed on Ly$\alpha$ narrow
band contours.
The I-band contains nearly pure continuum emission, with the relatively weak MgII 2798 line 
falling at one end of the filter band; the K-band is contaminated by
the H$\alpha$ emission line falling at the lower edge of the filter; finally the NICMOS H-band
contains a contribution from the [OIII] line (see Table 2).
\\
The emission from the ionized gas is very  extended ($\sim 9''$)
and is not associated with
the optical and near infrared emission;
the Ly$\alpha$ does not follow the optical/near
infrared components, but peaks at the location where the radio jet bends
sharply. This case is very
similar to that of MRC 1138$-$262 (Pentericci et al.\ 1997) and 1243$+$036 
(van Ojik et al.\ 1996): the most likely
interpretation is  that a cloud of ionized gas is responsible for
the deflection of the radio jet, and that the gas emission is
strongly enhanced in this region due to
 shock ionization.
Note also that the optical galaxy is located at a minimum of the Ly$\alpha$
emission, a common feature of many HzRGs (e.g. MRC 1138$-$262 and MRC 2104$-$242
below).
\\
Another interesting feature of MRC 0156$-$252 
is the presence of  two smaller components aligned (within 5$^{\circ}$) with 
the the radio axis but located {\it beyond} the radio hot-spots. They are  
of comparable brightness in the H-band (see Table \ref{tbl-4}), 
but the easternmost
clump, labeled C in Fig.\ 3, is much redder than clump B.
This difference in colors may be due to
a difference in age of the stellar population: 
McCarthy et al.\ 1992 argued on the basis of ground-based data that the
central galaxy contains a very old stellar populations 
($\sim 2.8$ Gyrs) or  alternatively a very reddened QSO spectrum. 
The bluer colors of component B, which is located near the bent in the radio jet,
might be interpreted as
enhanced star formation induced by the passage of the jet (e.g. Best et al.\ 1997) 
\\
\\
{\bf MRC 0406-244}
\\
This galaxy is  extensively discussed in Rush et al.\ (1997).
In Fig. 21a-c we show the NICMOS image in grey scale overlayed with contours
from (a) the I-band emission, (b) the K-band emission and (c) the Ly$\alpha$
narrow band emission.
In Fig. 21d we show in grey scale the radio jets at 8.2 GHz
superimposed to Ly$\alpha$
narrow band contours.
The K and H filters contain contributions from line emission:
in particular the K-band contain the strong H$\alpha +$[N II]
$\lambda\lambda$6548, 6584 system
and the H-band contains the [OIII] $\lambda\lambda$4959,5007 lines.
The I-band is purely continuum emission.
\\
The main body of the galaxy shows two emission peaks, and a number of
components  distributed in a sort of envelope on both sides.
The galaxy along the radio axis, which coincides remarkably well with a radio
peak, is most probably at a different redshift (Rush et al.\ 1997).
There is a spatial shift in the peak of the emission at different
wavelengths: the peak of the I-band coincides  with that of  H-band while
the K-band peak probably lies in between the two NICMOS peaks, although
due
 to the lower resolution (0.9$''$) and larger pixels size of this
image, with the present overlay we cannot exclude
that the K-band peaks coincides  with the brightest H-band peak.
\\
The  net color gradient between the northern side and the southern was
interpreted by Rush et al.\ (1997) with the fact that most of the emission
in the southeast extension is the result of line contamination.
The northwest components that appear in the I-band are probably true
continuum flux.
\\
A high resolution spectrum of the Ly$\alpha$ line taken along 
the direction of
the radio axis is presented in Fig. 22.
The spectrum has a total spatial extent of $\sim 8''$, similar to the 
extent seen in the narrow band image.
The emission comes from two components 
which are offset both in space ($\sim 2''$)
 and velocity ($\sim$1150 km sec$^{-1}$);
with the help of the narrow band image we identify
the brightest component with the main body of the galaxy,
while the other component is the northwestern extension.
They have  both a FWHM of $\sim 850$ km sec$^{-1}$.
An interesting feature is the deep trough which 
maybe due to absorption by associated neutral hydrogen, a
 feature common to many HzRGs
(van Ojik et al.\ 1997). The neutral gas is as spatially extended as 
the ionized gas and has a velocity gradient of $\sim 200$ km sec$^{-1}$.
\\
\\
{\bf MRC 2104-242}
\\
In Fig. 23a-c  we show the NICMOS image in grey scale overlayed with contours
from (a) the R-band emission, (b) the K-band emission and (c) the Ly$\alpha$
narrow band emission.\ Note the residuals of a spike
of a bright nearby star located to the north of the radio galaxy.
 We do not show an overlay of the Ly$\alpha$ gas with the radio emission
since the radio hot-spots are located further beyond the region of the
gas emission (see the radio map in Fig. 16). The radio 
axis is oriented at 12$^{\circ}$ (counterclockwise), implying that
both the continuum and line emission are well aligned with it.
In Fig. 24 we present a high resolution spectrum of the Ly$\alpha$ line.
\\
The galaxy is comprised of several components,
each having very different colors:
(i) the nucleus of the radio galaxy, which has a very red color $V-H \sim 5$, (ii) a series of
smaller bluer clumps northern of it; (iii) a
narrow elongated feature to the south.
\\
As in MRC 0156$-$262, the nucleus coincides with a minimum of  Ly$\alpha$
emission.
To the north of it, a bright component is  clearly seen in R-band and
is associated  with the brightest Ly$\alpha$ emission,
which at this position has a FWHM of $\sim$ 500 km sec$^{-1}$, a rather low
value compared with average HzRGs, but consistent with the
relation found by van Ojik (1995) that the largest radio sources tend
to have Ly$\alpha$ emission with lower velocity dispersion.
\\
Finally, the narrow elongated feature is associated with spectacular Ly$\alpha$ emission,
which  has 3 different peaks, having velocities of
140 km sec$^{-1}$, 670 km sec$^{-1}$ and 990 km sec$^{-1}$  blue-wards
of the central Ly$\alpha$ emission.
The overall velocity dispersion of the system is 1270 km sec$^{-1}$ (FWHM).
\\
Note that the Ly$\alpha$ spectrum shows that the Ly$\alpha$ emission extends
beyond that seen in the Ly$\alpha$ image and spans a total region 
of $\sim$14$''$, as large as the radio source.

\section{Discussion}
\subsection{Morphologies: when do the first elliptical galaxies appear?}
One of the most interesting problems in cosmology is at what epoch
elliptical galaxies formed.
To study this issue,
we tried to determine if any of the host galaxies of these distant radio
sources has a morphology that resembles that of elliptical
galaxies, and if so, determine their characteristic parameters
and their dependence with redshift.
\\
When possible, we fitted the radially averaged surface brightness profiles 
of the host galaxies with the most commonly used model distributions:
de Vaucouleurs and exponential. This was not done for (i) the 3
unresolved galaxies and (ii) the 2 galaxies in which the emission
in the central region is dominated by a point source. In these
last cases, although we could subtract a central point source revealing the
extended underlying galaxies, the subtraction was not good
enough to allow a fit of the residual.
\\
Whenever there were companions in the vicinity of the main galaxy to be fit, 
they where  subtracted and replaced by an average background value.
\\
For 9 out of 14 galaxies, the fit gave meaningless
results, in particular for those having  more than one peak of
emission within the main body of the galaxy (e.g. MRC 0406$-$244) and 
for those with an irregular morphology. 
Also some apparently regular galaxies (such as MRC 0943$-$242) are not well
 represented by any of the above laws.
For 5 of the galaxies 
(namely MRC 0152$-$209, USS 0211$-$122, USS 1707$+$105A,
MRC 2048$-$272 and MRC 2224$-$273), the surface brightness profile 
could be well represented by a de Vaucouleurs law, 
while the exponential profile gave worst results.
In Fig.\ 25 we show  the fit that we have obtained for
these radio galaxies, and in Table 5
 we report the  best fit parameters, r$_e$ and $\mu_e$,
that were determined by minimizing the $\chi^2$.\ We also report   
the values of the reduced chi-squared, which
indicates that the fit are quite good in all cases.
\\
The average effective radius for the 5 galaxies is r$_e$ =$ 5.4\pm 1.6$ kpc
(adopting $H_0=50$ km
sec$^{-1}$ Mpc$^{-1}$, and $q_0$=0.5 ), with a
median value of 5.3 kpc.
\\
The PSF of NICMOS might influence the result of the fit.\
It is difficult to recover accurately its shape since
the NIC 2 frames are very small and contain none or few point
sources; furthermore it is known that the PSF of  NICMOS varies
both spatially and with time (Colina and Rieke 1997).\ 
Only for USS 1707$+$105 we could construct a model of the PSF using 2 
non-saturated point sources near the radio
galaxy. We then de-convolved the frame of USS 1707$+$105 and
fitted again the image obtained: the resulting parameters $r_e$ and $\mu_e$
are, within the error, comparable to those obtained by fitting the 
original image.
So we are confident that, at least for the 3
largest galaxies (namely USS 0211$-$122, USS 1707$+$105A and MRC 0152$-$209), 
the PSF does not influence the results of the fit.
\\
Another possible source of confusion might be the presence of spatially 
extended line emission, expecially if this is distributed differently from the
continuum emission. However as we see from Table \ref{tbl-2}, line emission should be negligible in all cases except for MRC 2224$-$273, where it contributes  for $\sim 30$\% of the total H-band flux.
\\
We emphasize that while the good 
fit to the R$^{1/4}$ law suggests that these galaxies might be
morphologically ellipticals/bulges, a knowledge of the
spectral energy distribution is also needed to fully determine
their nature.
\\
It is interesting to notice that all the galaxies which can be well represented
by the de Vaucouleurs law, are in the lowest part of our redshift range (below
z$\sim 2.4$). This hints to a possible redshift evolution, although there
is not net dependence of morphology on redshift.
We can compare these results to what is obtained for lower 
redshift 3CR radio galaxies. It it well known
 that a large fraction ($\sim 80\%$) of the z $\sim 1$ radio galaxies
have profiles which are well fit by a de Vaucouleurs law 
(e.g.\ Best et al.\ 1998; McLure et al.\ 2000).
In particular  McLure et al. (2000) 
derived basic parameters for a sample of z $\sim$ 1 3CR radio galaxies  
from I-band WFPC2 images (using F814W or F785LP filters). These images 
have a resolution similar to ours; in addition we selected 
from their sample, a sub-group of objects such that the rest-frame 
emission sampled by the observations was as similar as possible 
to that of our 5 galaxies.
In practice we selected the 9 galaxies where the rest-frame range sampled was 
fully above the 4000 \AA\ break.\
These galaxies are at a median redshift of 0.75.\ 
Their average r$_e$ is 11.9$\pm$2.4 kpc, with a median value of 8.5 kpc. 
\\
Although the two samples are small, the indication is that the z $\sim 2$ 
radio galaxies are at least a factor of 2 
smaller than the lower redshift radio galaxies.
Only one high redshift galaxy (USS 1707$+$105A) 
has a radius comparable to its lower redshift counterparts.
For reference, brightest cluster ellipticals in the local universe have
an average effective radius of 32 kpc, and values vary from 10 kpc to several 
tens of kpc (e.g. Schombert 1988).  
\\
\\
The above  results indicate that a dynamically relaxed
and relatively old stellar population  is already in place in most of the
z $\sim$ 1 systems, while it becomes rarer at redshift
z $\sim 2$ and disappears at even higher redshift. 
A possible evolutionary scenario for the hosts of powerful radio
galaxies that can be inferred from these results
is the following:
at redshifts from 3.5 to $\sim 2.5$ most galaxies
show considerable substructure and clumpiness, suggesting strong 
interactions and mergers. At z $\sim 2$ some host galaxies 
appear morphologically relaxed;
between z $\sim 2$ and z $\sim 1$ almost
all systems (up to 80 \%) evolve into fully developed  elliptical galaxies  
and during this time
the characteristic scale length of the galaxies  increases
on average by a factor of 2.\ As argued by Best et al.\ (1998) these galaxies
will then continue to accumulate matter through mergers and gas infall:
if the mergers occurs homologously (e.g.\ Schombert 1987), a
merger of a large galaxy with a small systems gives as a result 
a remnant with a larger radius and a more diffuse morphology 
as compared to the original galaxy. 
In this way the  hosts of high redshift radio galaxies 
 would then evolve into  present day brightest cluster galaxies.

\subsection{Colors: comparison between WFPC2 and NICMOS morphologies}
For those objects (8) that have both the WFPC2 and NICMOS images, we
constructed a color index map in the following way.
The optical images were initially convolved with a Gaussian function
to bring them to the same resolution as the near infrared ones (0.2$''$).
For the galaxies USS 0211$-$122, MRC 0943$-$242, MRC
1138$-$242, USS 1410$-$001, USS 1707$+$105, MRC 2025$-$218 and MRC
2104$-$242 that were imaged with the planetary camera (Pentericci et
al.\ 1998,1999) we re-binned both frames on a finer pixel
scale; using factors of 3 and 5 for the WFPC2 and the NICMOS
respectively, the final pixel scales agreed to better than 1.3\%.
For the galaxy MRC 0406$-$242 which has been imaged on the WF3 chip
(Rush et al.\ 1997), the factors used were 4 and 3 respectively for
the WFPC2 and the NICMOS, and the agreement between the final
scales was better than 2\%.
\\
The WFPC2 and NICMOS images were aligned by comparing the location
of stars that were present on both frames. If this was not possible (e.g.
 because on the small NICMOS frame there were no point sources) we used
the location of distinct peaks of the radio galaxy itself. We only
applied integer shifts to avoid interpolation. After sky
subtraction we calibrated both images to units of flux density in
$\mu$Jy, using the information in the headers of the WFPC2 images
and the  calibration of the NICMOS Camera 2 filters derived by
Colina and Rieke (1997).
\\
Finally we
derived color maps as: \\
COLOR $= log(S_{\it WFPC2}) - log(S_{\it NICMOS})$, \\
excluding all pixels
having fluxes less than 3$\sigma$ (where $\sigma$ is the rms
noise), and re-binned the final image by
averaging regions of 2x2 pixels. The resulting image can be
calibrated on a scale of two-point spectral index $\alpha$, where
$S_{\nu} \sim \nu^{- \alpha}$, or alternatively can be scaled to
magnitude color index as: (i) R-H=1.17+0.91$\alpha$, for those galaxies
that were imaged with the F702W filter (all galaxies at
redshift beyond 2.9), or (ii)
V-H=1.40+1.08$\alpha$ for
those that were imaged with the F606W filter (all other galaxies).
\\
The resulting maps of two-point spectral index are presented in Figs. 26-29.
Note that there are some spurious color structures,
like sharp pixel to pixel variations,
especially at the edges of some small components, due to regions
with very different signal to noise in the 2 colors,
the  difference in PSF shape, and the uncertainty of $\sim 0.1''$
in the alignment procedure.
\\
We shall use these color maps in the next section to
study the alignment effect.
Here we just examine the distribution of the color indices and
their possible relations to other properties of the radio sources.
\\
We limit this study
to the clumpiest objects, which have a complex color index distribution, i.e.\
the four galaxies MRC 0406$-$244, MRC
1138$-$262,  USS 1707$+$105 and MRC 2104$-$242.
We measured the color index in different regions
of the above galaxies by averaging over a circle with 
$\sim 0.2''$ radius around the peaks in the emission present in
 the WFPC2 image and/or in the NICMOS image (some components are
 present only in one of the two bands). We only considered
regions of emission that are surely part of the systems i.e. that
emit Ly$\alpha$ at the same redshift and/or are embedded in the
Ly$\alpha$ halo of the galaxies. In Fig. 30  we present an
histogram with the distribution of optical to infrared spectral
indices. The median color index is around $\alpha=2$ and more than
75\% of the components have spectral indices between 1.5 and 3
(corresponding to V-H colors between 3 and 4.6), indicating that
the radio galaxies contain very red regions.
\\
Note that since the redder bands is in some cases
(e.g. MRC 2104$-$242, see Table 2) contaminated  by line emission,
a very irregular and clumpy distribution of the emitting gas could
produce sharp color gradients
between the various components.
\\
Since many properties of HzRGs depend on radio size (e.g.\ van Ojik
et al.\ 1997) it is interesting to determine if there is a correlation
between the  color index of the clumps and their distance from the
radio core. The results are show in Fig. 31: there's no
evident relation between the distance of the component and its   
color. We also looked at a possible correlation between the color
index of a component and the relative position angle of the
component with respect to the radio axis: again we see no
correlation.
Therefore we conclude that the colors of the aligned components of
HzRGs, don't vary much with radio size and orientation, although
presenting a large scatter. Note however that the statistics is
small and that there is not a large variation in size between the
4 galaxies considered, so this could mask possible relations.

\subsection{The radio/near-infrared alignment effect}
Our images show that the near-infrared continuum emission of HzRGs
is generally aligned with the main axis of the radio emission.
This is the so-called alignment effect which has been studied for
more than a decade. Several models have been proposed to explain
its nature (for a review see  McCarthy 1993a and references
therein), the most viable ones being: (i) star-formation
stimulated by the radio jet as it propagates outward from the
nucleus (Chambers et al.\ 1987; McCarthy et al.\ 1987; de Young
1989; Daly 1990); (ii) scattering of light from an obscured
nucleus by dust or free electrons (di Serego Alighieri et al.\
1989; Scarrot et al.\ 1990; Tadhunter et al.\ 1992; Cimatti et
al.\ 1993); (iii) nebular continuum emission from warm line
emitting clouds (Dickson et al.\ 1995). There is evidence that all
these mechanisms contribute to the alignment effect, with strong
variations from object to object (e.g. Pentericci et al.\ 1999).
We will try to asses their contribution using the new
near-infrared images and the color maps constructed in the
previous section.
\\
Assuming that a considerable fraction of the blue light is scattered light
from a central quasar 
(see for example Cimatti et al.\ 1994), we can estimate
the percentage of scattered light there would be in the H-band.
According to unification models the incident spectrum seen
by the scattering medium is that of a quasar,
$S\propto\nu^{-\beta}$ (we use $\beta$ to distinguish this 
parameter from the optical-to-infrared spectral index, defined as $\alpha$) . 
As a value for $\beta$ we take the one derived from
the composite spectra of quasars from the MRC catalogue
(the same parent catalogue from which most of our sources were taken) compiled
by Baker and Hunstead (1995). They found $\beta =1$ for the steeper cases.
Thompson scattering by thermal electrons is wavelength independent
(although in some cases it has serious energetic difficulties,
e.g. Eales and Rawlings 1990): in this case
the scattered spectrum will be in shape similar to the incident one.
For dust scattering, the output spectrum will be bluer than the incident one
(for optically thin dust), with the exact shape depending on the size of
the grains
(e.g.\ Dey et al.\ 1996 and references therein).
\\
Assuming 50\% of the blue flux (at $\sim 1800$ \AA \space
rest-frame) is scattered light, and that the scattered component
has a  power-law spectrum with $\beta =1$, we can calculate its
contribution at $\sim 5000$ \AA \space for components with
different observed  optical-to-infrared spectral indices. The
fraction of scattered light in the optical will also be 50\% for
those components with an observed spectral index $\alpha=1$; it will be 26\%
for the components with an observed optical-to-infrared spectral
index  $\alpha=2$; it will be 10\% for those with an index
$\alpha=3$; and finally it will be 4\% for the components with
$\alpha=4$. These percentages would decrease considerably if the
scattering is wavelength dependent; on the other hand they could
increase if we allow for some reddening. However it is clear that
for components with observed optical-to-infrared spectra steeper
than $\alpha\sim 2$, as most radio galaxies exhibit (see Fig. 30),
scattered light cannot be the dominant fraction of the emission at
5000 \AA \space rest-frame wavelength.
\\
\\
We note that in some objects
the aligned component has a nearly uniform spectral index
whereas in other ones, whereas in other ones.
To the first category belong the galaxies  USS 0211$-$122, MRC 0943$-$242
USS 1410$+$001 and MRC 2025$-$218 (although this last case is more
complicated due to
the presence of a strong unresolved nuclear component), where
the aligned light has a nearly constant spectral index ($\sim 2.3$ for
USS 0211$-$122 and MRC 2025$-$218, and $\sim 2$  for MRC 0943$-$242 and
USS 1410$+$001).
\\
For two of these objects there exist also polarization measurements,
showing  that the rest-frame UV emission is polarized: USS 1410$-$001 has
a polarization of 10\% (Cimatti et al.\ 1998)
and an estimated total contribution of scattered light of 40 to
60\% (depending on the nature of the scatterer) at 1800 \AA~ (rest-frame).
MRC 2025$-$218 has a polarization of 8\% (Cimatti et al.\ 1994), and the total
SED of the galaxy can be fit by models in which scattered light
accounts for a very large fraction of the emission at 1800 \AA.
\\
Then from the uniformity of the spectral indices in the aligned component,
and  using the calculations above,
we conclude that in these objects also a considerable fraction
($\sim 30-40$\% or more)
of the rest-frame optical emission could  be scattered light.
\\
\\
For the other galaxies (namely MRC 0406$-$244, MRC 1138$-$262, USS 1707+105 and MRC
2104$-$242) 
the optical-to-infrared color index varies considerably
within the different regions.
Furthermore, the spectral indices of some components are considerably
steeper than $\alpha=2$, and
 in some galaxies the components are
located outside the
ionization cone. In the source MRC 1138$-$262 if we were to draw such a cone 
so that all the continuum emission would line inside, 
it should have an half opening angle of at least 70$^{\circ}$.
This is much wider than what expected by models of AGN unification 
(up to 45$^{\circ}$ , e.g. Barthel 1989), and than values found through 
imaging of the  the emission line regions in active galaxies: for example 
Wilson and Tsvetanov (1994) find average half opening angles 
of $\sim$ 30$^{\circ}$.
The same thing is valid for USS 1707+105A,
which is clearly extended in a direction perpendicular to the radio source
and therefore outside the ionization cone.
These characteristics argue against  scattering as a major contributor to
the light, since in this case we would expect components with uniform and
bluer colors (assuming that 
the properties of the scattering medium do not vary considerably 
amongst the various components).
Hence we conclude that in these systems 
the contribution of scattered light to the optical continuum 
must be very small (less than 10\%).
\\
If we  then assume that in these galaxies most of
the emission is produced stellar light, we can derive an approximate age for 
the stellar population and compare it with the age of the radio sources, to
determine  whether jet-induced star formation could be a viable mechanism 
to account for the stellar population.\
To derive an {\it indicative} age for the stars, we have matched the
amplitude of the 4000 \AA \space break observed in the clumps with
the amplitudes (at the same rest-frame wavelengths) derived from
the spectral energy distributions of stellar populations evolving
according to the galaxy isochrone synthesis spectral evolution
library compiled by  Bruzual and Charlot (1993).\ We used their 1995
model, assuming a simple stellar population, and a Salpeter
initial mass function (IMF), with star masses ranging from 0.1 to
125 $M_{\odot}$. The instantaneous star-burst model, which 
evolves most rapidly, can reproduce the observed 4000  \AA
\space break in 0.5 Gyrs for the bluest components of HzRGs, to
1.3 Gyr for the reddest ones. All other modes of star formation
(e.g. star-burst with a finite duration, an exponentially
declining star formation mode, etc.)
 require a longer time to reproduce the same colors.
\\
Clearly there are a number of uncertainties in this age
determination, e.g. the inferred ages could be wrongly estimated in
those components that have a contribution of line emission in any
of the two bands. In any case the indication is that these
components have rather large ages, of the order of half a Gyr or
more. 
\\
The typical 
life-time of a radio source can be easily derived as t$=$D/2v,
where v is the expansion speed of the hot-spots and D is
the total extent of a radio source. The velocity v is typically in
the range  0.01-0.2c (e.g.\ Alexander and Leahy 1987), 
and our sources have maximum extent of about
200 kpc. Therefore their lifetime will be in the range 0.2-3 $\times 10^7$
yrs, i.e. about 100 times
shorter than that of the stellar population. 
Note that alternative estimates of the radio source life-time, e.g. 
from spectral ageing arguments also in general lead
to ages of $\sim 10^7$ yrs (Carilli et al.\ 1991).
Clearly if the radio activity was
recurrent then the observed clumps could have formed during
previous phases of activity.
\\
Finally we mention the model proposed by West (1994), in which the
anisotropy of
the optical and infrared emission around a typical high redshift
radio galaxy is due to the anisotropy of the surrounding density
distribution (see also Eales 1992). In this model, galaxy formation
proceeds along preferred directions which are related to large
scale elongated distribution of matter: in the prolate potential
of the forming galaxy, the gas will fall and settle into a disk whose
axis is along the
major axis of the distribution of matter. The radio jets then will
be emitted along this axis, hence the alignment between their
direction and the distribution of the surrounding  material (gas
and stars). Within this model the relative ages of the radio jets
and stellar components do not  have to be matched. However, this
model alone cannot explain the alignment effect since, for
example, it cannot explain the polarization properties of many
HzRGs.

\subsection{Companion galaxies and their relation to the alignment effect}
 Most HzRGs in our sample show very close companion objects.
We define a companion as a small (size  $\le 1''$) object that is
located within a region delineated by a 100 kpc diameter around the
radio source. We exclude objects with a stellar profile (although
they might be in some cases high redshift unresolved objects,
rather than stars) and those which can be recognized as spirals,
on the assumption that they are most probably foreground systems.
We do not consider the inner 30 kpc from the  
galaxy centers, to ensure that the regions of the 
source itself are excluded from the statistics 
(this is true in almost all cases except for the larger 
systems, but the overall result does not depend on them).    
\\
The average number of companion galaxies for the 19 HzRGs is
1.9$\pm$0.3 down to a magnitude limit of $m_H=23.5$. From deep counts of
NICMOS parallel orbit,
Yan et al.\ (1998) derive an average cumulative density of galaxies
in H-band down to a magnitude $m_H=23.5$ of $\sim 1.5\times 10^5$ galaxies
deg$^{-2}$. The expected number of galaxies in a circular area of
radius 50 kpc is 1.4. Therefore there is a marginal (2$\sigma$) 
excess of objects in the area around HzRGs.
\\
We do not have any other systematic information on these possible 
companion galaxies, that might tell us about their real redshifts. 
However further evidence that these objects may be associated  to the radio
galaxies is provided by the distribution of the orientation of
companions with respect to the radio axis (or optical axis if the
radio source is unresolved). In computing this distribution we
have excluded companions of the 2 radio galaxies that are unresolved
both in the radio and in the near-infrared emission (MRC 0324$-$228 and 
MRC 1017$-$220)
since in these cases it is impossible to
determine any preferred direction. 
The result is presented in Fig.\ 32: the
histogram shows that the companion galaxies are predominantly
located along the radio axis. A Kolmogorov-Smirnov test shows that
the distribution differs from a uniform distribution between
0$^{\circ}$ and 90$^{\circ}$ with 99\% probability.
\\
R\"ottgering et al.\ (1996) reported a similar effect, finding a 
statistical excess
of optical companions located along the radio axis of USS radio sources. 
While they defined companions as objects
that are located within a circle with diameter equal to the radio
source extension, we prefer to use a fixed scale, since we do not assume 
a-priori that the companion galaxies are related to the presence
of the radio sources. 
However it is
significant that, even with a different definition,
both groups find an excess density in the direction of the
radio axis.
\\
This result shows that the alignment effect not only holds
for the radio galaxy hosts, but also for its neighboring galaxies 
within an area of at least $\sim 100$ kpc.
An interesting issue would be also to determine how the overdensity and 
the alignment of these companion galaxies  
depend on the area considered. Unfortunately the NICMOS data are 
not useful in this respect, since the NIC 2 field of view 
is far too small. 
\\
The above conclusions show that 
any model attempting to explain the alignment effect has to take
into account not only the morphology of the radio galaxies hosts,
but also their surroundings 
(for the possible origin of the alignment effect see 
the discussion and the references in the previous section).

\section{Conclusion}
We have presented the results of a near infrared imaging program
on a sample of 19 radio galaxies having redshift between 1.7 and
3.2, observed with NICMOS Camera 2 on the Hubble Space Telescope.
\\
Our observations show that the host galaxies of powerful radio sources
have  a wide range of morphologies,
from systems with simple compact morphologies to systems
with substructures such as multiple emission peaks  to systems
comprised of several components spread over large areas.
Only in few cases the active nucleus
dominates the emission in the central region.
\\
Following is a summary of the most important results obtained:
\\
(i) While most systems appear irregular, 5 galaxies 
at z $\sim$ 2 have  morphologies consistent with them being elliptical
galaxies (or bulges), i.e. they can be well represented by the classical R$^{1/4}$ law.
A comparison between this small sample and the host galaxies of 
z $\sim 1$  3CR radio sources, observed at similar rest-frame wavelengths and resolution,
indicates 
 that there is  difference of a factor 2 in their effective radii.
\\
 (ii) We find that in  almost all extended systems the near infrared continuum
light is  aligned with the radio axis and the aligned component has 
very red colors. For several galaxies
WFPC2 V or R-band images were available so we computed a high
resolution map of the optical-to-infrared spectral index. These spectral indices are in general
rather steep, $\alpha \sim 2$ or steeper.
The clumpy systems show net color differences between the various components, 
which  can be explained with different ages for the
stellar content.
Indicative ages determined from the Bruzual and Charlot models
are of the order of $\sim$0.5-1 Gyr in most cases, much larger than
typical radio source ages.
\\
(iii) In the small NICMOS frames we can see in many cases nearby
faint objects around the high redshift radio galaxies, and their
space density is slightly higher than what is observed in the
deep NICMOS parallel observations of random fields. Furthermore 
these objects tend to be aligned with the direction of the
main axis of the radio sources suggesting that they are related to
the presence of the radio galaxies.




\acknowledgments

This work is based on observations with the NASA/ESA Hubble Space Telescope,
obtained at the Space Telescope Science Institute, which is operated by AURA
Inc.\ under contract with NASA. 
HJAR acknowledges support from an EU
twinning project, a programme subsidy granted by the Netherlands
Organization for Scientific Research (NWO) and a NATO research grant.
The work by WvB at IGPP/LLNL was performed under the auspices of the 
US Department of Energy under contract W-7405-ENG-48.



\clearpage

\begin{deluxetable}{lrrrrcrc}
\footnotesize \tablecaption{Properties of the radio sources.
\label{tbl-1}} \tablewidth{0pt} \tablehead{ \colhead{Source} &
\colhead{z}   & \colhead{RA}   & \colhead{Decl} &
\colhead{S$_{4.7}$}  & \colhead{$\alpha_{8.2}^{4.7}$} &
\colhead{Size} & \colhead{Morphology} \\
\colhead{(1)} & \colhead{(2)} & \colhead{(3)} & \colhead{(4)} & 
\colhead{(5)} & 
\colhead{(6)} & \colhead{(7)} &\colhead{(8)}  
 } 
\startdata 
MRC 0140$-$257&2.64& 01h42m41.16s&-25d30m34.1s & 47  &1.3 & 4.2 &double\\ 
MRC 0152$-$209&1.89& 01h54m55.77s&-20d40m26.3s & 115 &1.9 & 1.6 &one-sided\\ 
MRC 0156$-$252&2.09& 01h58m33.45s&-24d59m30.2s &112&1.1&8.3&double,distorted\\
USS\hskip0.25cm 0211$-$122&2.34& 02h14m17.37s&-11d58m46.7s & 54 &1.5 & 17.0&double \\
MRC 0316$-$257&3.13& 03h18m12.06s&-25d35m09.7s & 101&1.3 & 7.6 &double\\ 
MRC 0324$-$228&1.89& 03h27m04.44s&-22d39m42.6s & 131&1.2$^{\it a}$& 9.7& double, no core \\ 
MRC 0350$-$279&1.90& 03h52m51.64s&-27d49m22.6s &86  &1.2$^{\it a}$&$\le$0.6&unresolved\\ 
MRC 0406$-$244&2.44& 04h08m51.44s&-24d18m16.7s & 108 &1.3 & 10.0&double\\
MRC 0943$-$242&2.93& 09h45m32.79s&-24d28m49.8s & 55  &1.8&3.9&double,no core\\
MRC 1017$-$220&1.77& 10h19m49.05s&-22d19m58.0s&261&1.1 &$\le$0.6&unresolved\\ 
MRC 1138$-$262&2.16&11h40m48.25s&-26d29m10.1s&153&1.8&15.8&double,distorted\\ 
USS\hskip0.25cm 1410$-$001&2.33& 14h13m15.13s&-00d22m59.6s & 57  &1.3 & 24.0&double\\
USS\hskip0.25cm 1707+105&2.35& 17h10m06.85s&+10d31m09.0s &64&1.2&22.5&double,no core\\ 
MRC 2025$-$218&2.63& 20h27m59.45s&-21d40m57.1s& 95 &1.1&5.1&double,distorted\\
MRC 2048$-$272&2.06& 20h51m03.37s&-27d03m04.6s & 115 &1.6 & 8.3 &double\\
MRC 2104$-$242&2.49& 21h06m58.16s&-24d05m11.3s & 107 &1.0 & 24  &double\\ 
USS\hskip0.25cm 2202+128&2.70& 22h05m14.27s&+13d05m33.7s & 52  &1.5 & 4.2 & double\\
MRC 2224$-$273&1.68& 22h27m43.26s&-27d05m01.7s &60 &1.6&$\le$0.6&unresolved\\ 
USS\hskip0.25cm 2349+280&2.89& 23h51m59.08s&+29d10m28.9s & 34 &1.6$^{\it b}$&15.2& double \\

\enddata

\tablenotetext{a}{ Spectral index between 4.7 Ghz and 408 MHz}

\tablenotetext{b}{Spectral index between 4.7 GHz and 1.4 GHz}

\tablecomments{(1) 1950IAU name of the radio galaxy; (2) redshift; (3) and
(4) position  of the radio core (J2000 coordinates); (5) total 
flux and 4.7 GHz; (6) total spectral index between 4.7 and 8.2 GHz;
(7) total extension in arcsecond; (8) radio morphology at 8.2 GHz  
}

\end{deluxetable}

\clearpage

\begin{deluxetable}{lrrrrrrcr}

\footnotesize \tablecaption{NICMOS observations .
\label{tbl-2}} \tablewidth{0pt} \tablehead{ \colhead{Source} &
\colhead{Date}   & \colhead{ N}   & \colhead{$\mu_{(3\sigma)}$} &
\colhead{T$_{exp}$} & \colhead{Filter} & \colhead{$\lambda_r$} &
\colhead{Cont. lines} & \colhead{\%flux} 
\\
\colhead{(1)} & \colhead{(2)} & \colhead{(3)} & \colhead{(4)} & 
\colhead{(5)} & 
\colhead{(6)} & \colhead{(7)} &\colhead{(8)} &\colhead{(9)}
 } 
\startdata 
MRC 0140$-$257& 17/10/97&4& 25.00&10195& F160W&3850-4950&H$\beta$ &1 \\ 
MRC 0152$-$209& 26/12/97&4& 24.48&10259& F160W&4840-6230&[OIII] &6 \\
MRC 0156$-$252& 28/09/97&4& 24.89& 9232& F160W&4530-5830&[OIII],H$\beta$&5\\
USS\hskip0.25cm 0211$-$122& 19/10/97&4& 24.65&10195& F160W&4190-5390&[OIII],H$\beta$&3\\
MRC 0316$-$257& 17/10/97&4& 24.76&10195& F160W&3390-4360&[OII] &10 \\
MRC 0324$-$228& 22/06/97&2& 24.50& 5065& F165M&5540-6230& --&-- \\
MRC 0350$-$279& 18/10/97&4& 24.63&10195& F160W&4830-6210& [OIII] &-- \\ 
MRC 0406$-$244& 11/08/97&5& 24.84&12824&F160W&4070-5230&[OIII],H$\beta$&17\\
MRC 0943$-$242& 12/06/97&3& 24.55& 7694& F160W&3560-4580& [OII] &7\\
MRC 1017$-$220& 13/06/97&2& 24.25& 5065& F165M&5780-6500& [OI] &--\\
MRC 1138$-$262& 27/11/97&5& 24.57&12824&F160W&4430-5700&[OIII],H$\beta$& 2\\
USS\hskip0.25cm 1410$-$001& 07/01/98 &4& 24.76&10195&F160W&4200-5410&[OIII],H$\beta$&9\\
USS\hskip0.25cm 1707+105& 24/10/97 &4& 25.01&10195& F160W&4180-5370&[OIII],H$\beta$&2\\
MRC 2025$-$218& 09/09/97&4& 24.64&10259& F160W&3860-4960& H$\beta$& 1\\
MRC 2048$-$272& 23/10/97&4& 24.65&10259&F160W&4580-5880&[OIII],H$\beta$&--\\
MRC 2104$-$242& 24/10/97&4& 24.61&10259&F160W&4010-5160&[OIII],H$\beta$&26\\
USS\hskip0.25cm 2202+128& 21/06/97 &2& 24.25& 5065& F165M&4320-4860& H$\beta$ &2\\
MRC 2224$-$273&11/11/97&4& 24.68&10259& F160W&5220-6720& H$\alpha$,[OI]&28\\
USS\hskip0.25cm 2349+280& 24/11/97 &4& 24.78&10195& F160W&3600-4630& [OII]&2 \\

\enddata

\tablecomments{(1) 1950IAU name of the radio galaxy; (2) observation date; (3) number of orbits;(4) limiting surface brightness in H magnitudes; (5) total exposure time (seconds); (6) filter used; (7) restframe wavelegth range 
covered by the observations; (8) emission lines  falling in the continuum band;(9) estimated percentage of line flux in the total emission measured 
within a 4$''$ circular aperture.
}

\end{deluxetable}

\begin{deluxetable}{lrrccrcr}
\footnotesize \tablecaption{Ground based observations log
\label{tbl-3}} \tablewidth{0pt} \tablehead{ \colhead{Galaxy} &
\colhead{Telescope}  & \colhead{Date}   & \colhead{Band} &
\colhead{$\lambda_0$} &\colhead{T$_{exp}$}  &  \colhead{Res.} &
\colhead{Ref} 
\\
\colhead{(1)} & \colhead{(2)} & \colhead{(3)} & \colhead{(4)} & 
\colhead{(5)} & \colhead{(6)} & \colhead{(7)}  &\colhead{(8)}
} 

\startdata MRC 0156$-$252 &du Pont 2.5m &27/10/89&

Ly$\alpha$& 3700/50 & 8100 &1.5$''$&1\\

         &du Pont 2.5m &15/11/96&  K$_s$       &  2.2$\mu$ & 4770 &0.9$''$&1\\

         &du Pont 2.5m &9/09/91 &  I           &  8100     & 1800 &0.9$''$&1\\

\hline

MRC 0406$-$244 & du Pont 2.5m &17/02/96& Ly$\alpha$  &4200/100&9000&1.1$''$& 2
\\

         & du Pont 2.5m &19/11/96& K$_s$       &  2.2$\mu$ & 9030 &0.9$''$&2\\

         & du Pont 2.5m & 8/09/91& I           &  8100     & 1300 &0.9$''$&2\\

         & NTT/EMMI     &24/11/97&Spectr. Grat 3&  4180    & 4800  &2.8\AA&1\\

\hline

MRC 2104$-$242 & du Pont 2.5m &29/10/89 & Ly$\alpha$ & 4265/50&7800&1.4$''$&1\\
         & du Pont 2.5m & 9/09/91 & $r$          & 6400    & 3100&0.7$''$ &1\\

         & NTT/EMMI     &30/06/97 &Spectr. Grat 3& 4240     &7200&2.8\AA &1\\

\enddata

\tablecomments{(1) Radio galaxy; (2) Telescope used; (3) observation date ;
(4) filter or grating used; (5) observed wavelegth covered (in \AA); (6) total exposure time (in seconds); (7) resolution; (8) reference 1. This paper; 2. Rush et al. 1997}

\end{deluxetable}
\clearpage

\begin{deluxetable}{llccc}

\footnotesize \tablecaption{Photometry
 \label{tbl-4}}
\tablewidth{0pt} \tablehead{ \colhead{Galaxy} & \colhead{Comp.} & \colhead{Ap.} &\colhead{$M_H$}  & \colhead{err}  } \startdata

  MRC 0140$-$257             & main & 4$''$ & 20.13 & 0.07\\
  MRC 0152$-$209             & main & 4$''$ & 18.69 & 0.04 \\
  MRC 0156$-$252             & A    & 4$''$ & 18.39 & 0.10\\
                             & B    & 2$''$ & 21.58 & 0.10\\
                             & C    & 2$''$ & 20.22 & 0.10\\
  USS\hskip0.25cm 0211$-$122 & main & 4$''$ & 19.64 & 0.04 \\
  MRC 0316$-$257             & main & 4$''$ & 20.23 & 0.05 \\
  MRC 0324$-$228             & main & 4$''$ & 19.93 & 0.03 \\
  MRC 0350$-$279             & main & 4$''$ & 20.26 & 0.10 \\
  MRC 0406$-$244             & A    & 4$''$ & 18.91 & 0.04\\
                             & B    & 1$''$ & 21.36 & 0.10 \\
                             & C    & 1$''$ & 21.23 & 0.10 \\
                             & D    & 1$''$ & 21.62 & 0.10 \\
                             & E    & 1$''$ & 22.67 & 0.10 \\
  MRC 0943$-$242             & main & 4$''$ & 19.77 & 0.08 \\
  MRC 1017$-$220             & main & 4$''$ & 18.28 & 0.05 \\
  MRC 1138$-$262             & A    & 4$''$ & 18.04 & 0.03 \\
                             & B    & 1$''$ & 23.13 & 0.20 \\
                             & C1   & 1$''$ & 21.11 & 0.10 \\
                             & C2   & 1$''$ & 22.07 & 0.10 \\
                             & E    & 1$''$ & 23.48 & 0.20 \\
                             & F    & 1$''$ & 22.04 & 0.10 \\
                             & H    & 1$''$ & 21.68 & 0.10 \\
  USS\hskip0.25cm 1410$-$001 & main & 4$''$ & 19.25 & 0.05 \\
  USS\hskip0.25cm 1707+105   &  A   & 4$''$ & 20.15 & 0.04 \\
                             &  B   & 1$''$ & 23.50 & 0.20 \\
                             &  C   & 1$''$ & 22.30 & 0.10 \\
                             &  D   & 1$''$ & 22.65 & 0.10 \\
                             &  E   & 1$''$ & 22.63 & 0.10 \\
  MRC 2025$-$218             & main & 4$''$ & 18.94 & 0.04 \\
  MRC 2048$-$272             & main & 4$''$ & 20.18 & 0.04 \\
  MRC 2104$-$242             & A    & 4$''$ & 19.91 & 0.10 \\
                             & B    & 2$''$ & 21.18 & 0.10 \\
  USS\hskip0.25cm 2202+128   & main & 4$''$ & 19.80 & 0.10 \\
  MRC 2224$-$273             & main & 4$''$ & 19.23 & 0.05 \\
  USS\hskip0.25cm 2349+280   & main & 4$''$ & 19.94 & 0.05 \\
  \enddata
\end{deluxetable}           

\clearpage
\begin{deluxetable}{lrcccc}

\footnotesize \tablecaption{Parameters of the de Vaucouleurs fit
\label{tbl-5}} \tablewidth{0pt} \tablehead{ \colhead{Galaxy} &
\colhead{z}  & \colhead{$r_e$}  & \colhead{$r_e$}  & \colhead{$\mu_e$}& 
\colhead{$\chi^2/\nu$}  \\
             &              & \colhead{arcsec} &  \colhead{kpc} & \colhead{H-mag arcsec$^{-2}$}  & }
\startdata

MRC 0152$-$209             & 1.89 & 0.64 & 5.3  & 20.8 & 1.33  \\
USS\hskip0.25cm 0211$-$122 & 2.34 & 0.73 & 5.7  & 21.8 & 1.11  \\
MRC 1707+105               & 2.35 & 1.62 & 12.7 & 23.7 & 1.06  \\
MRC 2048$-$272             & 2.06 & 0.2  & 1.6  & 19.1 & 0.64  \\
MRC 2224$-$273             & 1.68 & 0.2  & 1.7  & 18.6 & 1.12  \\
\enddata

\tablecomments{(1) Radio galaxy; (2) redshift; (3) best fitting r$_e$ in arcseconds; (4)
best fitting r$_e$ in kpc; (5) best fitting $\mu_e$; (6) reduced chi-squared  }
\end{deluxetable}

\clearpage

 \begin{figure}
\plotone{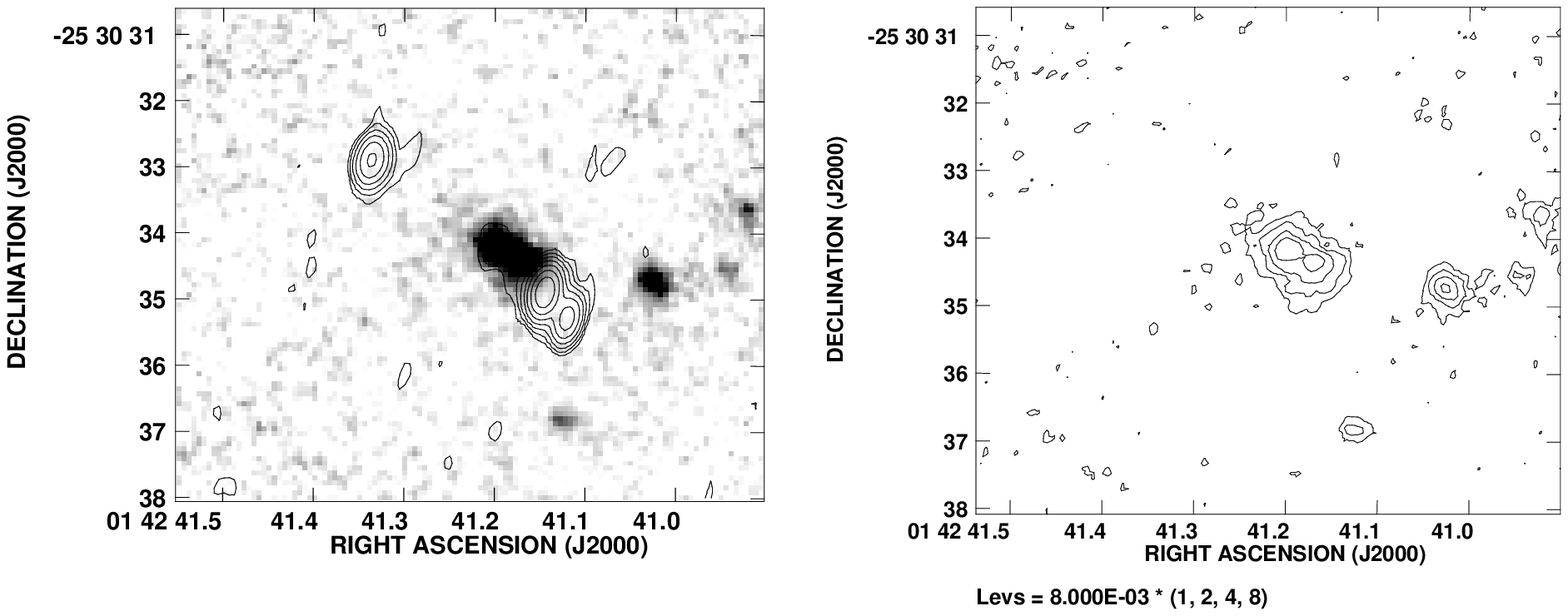}     
\caption[f1.ps]{{\it Left}: A grey scale representation of 
the near infrared 
continuum emission of the radio galaxy MRC 0140$-$257 at z $=2.64$
with contours  from the VLA 8.2 GHz observations super-imposed. 
{\it Right}: Contour representation of the continuum emission. Contour 
levels of flux density at: 1,2,4,8 $\times$0.8$\cdot$10$^{-2}$ $\mu$Jy.
 \label{f:0140}}
\end{figure}  
\begin{figure}
\plotone{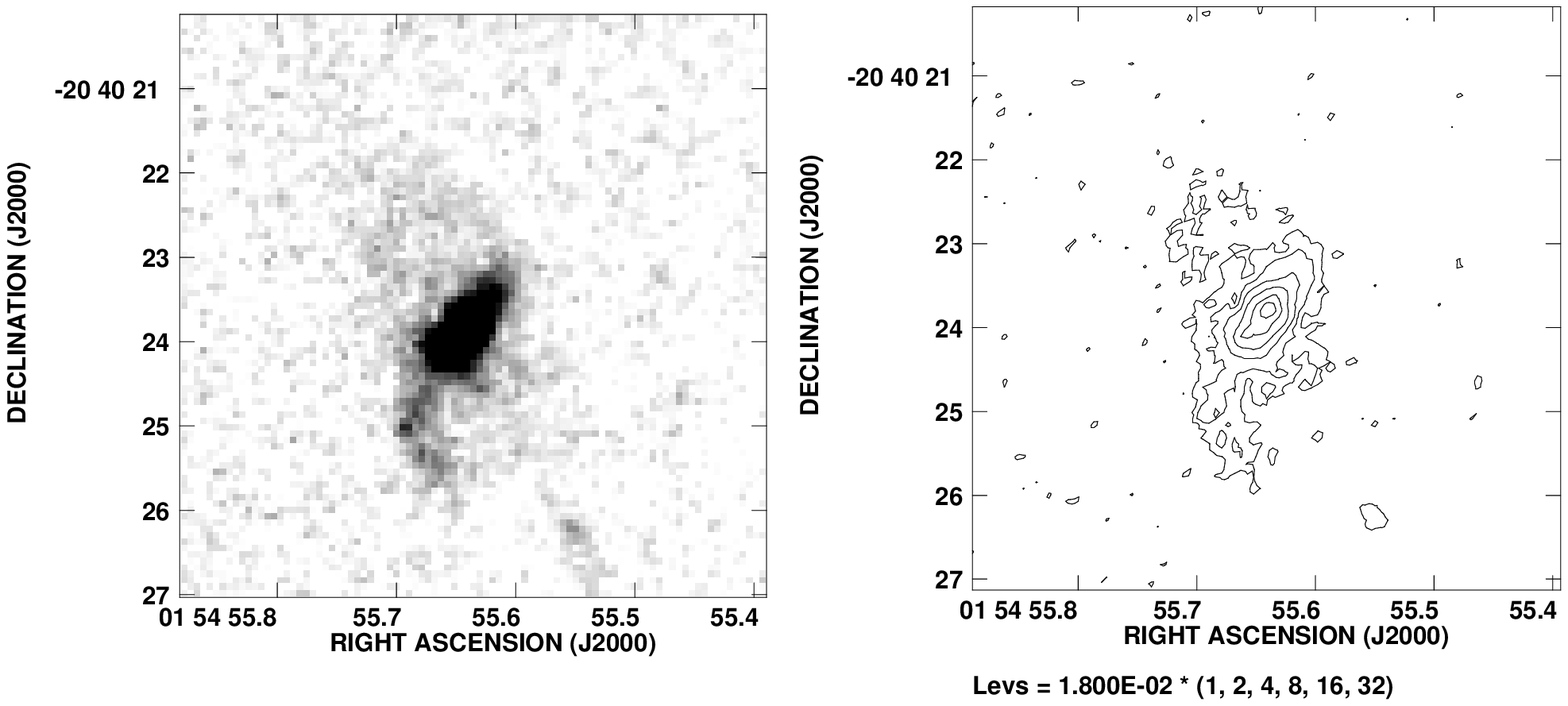}   
\caption[f2.ps]{{\it Left}: A grey scale representation of the near 
infrared continuum
emission of the radio galaxy
MRC 0152$-$209 at z $=1.89$. {\it Right}: Contour representation of the same
image.  Contour 
levels of flux density at: 1,2,4,8,16,32 $\times$1.8$\cdot$10$^{-1}$ $\mu$Jy.
{\label{f:0152}}}
\end{figure}  

\begin{figure}
\plotone{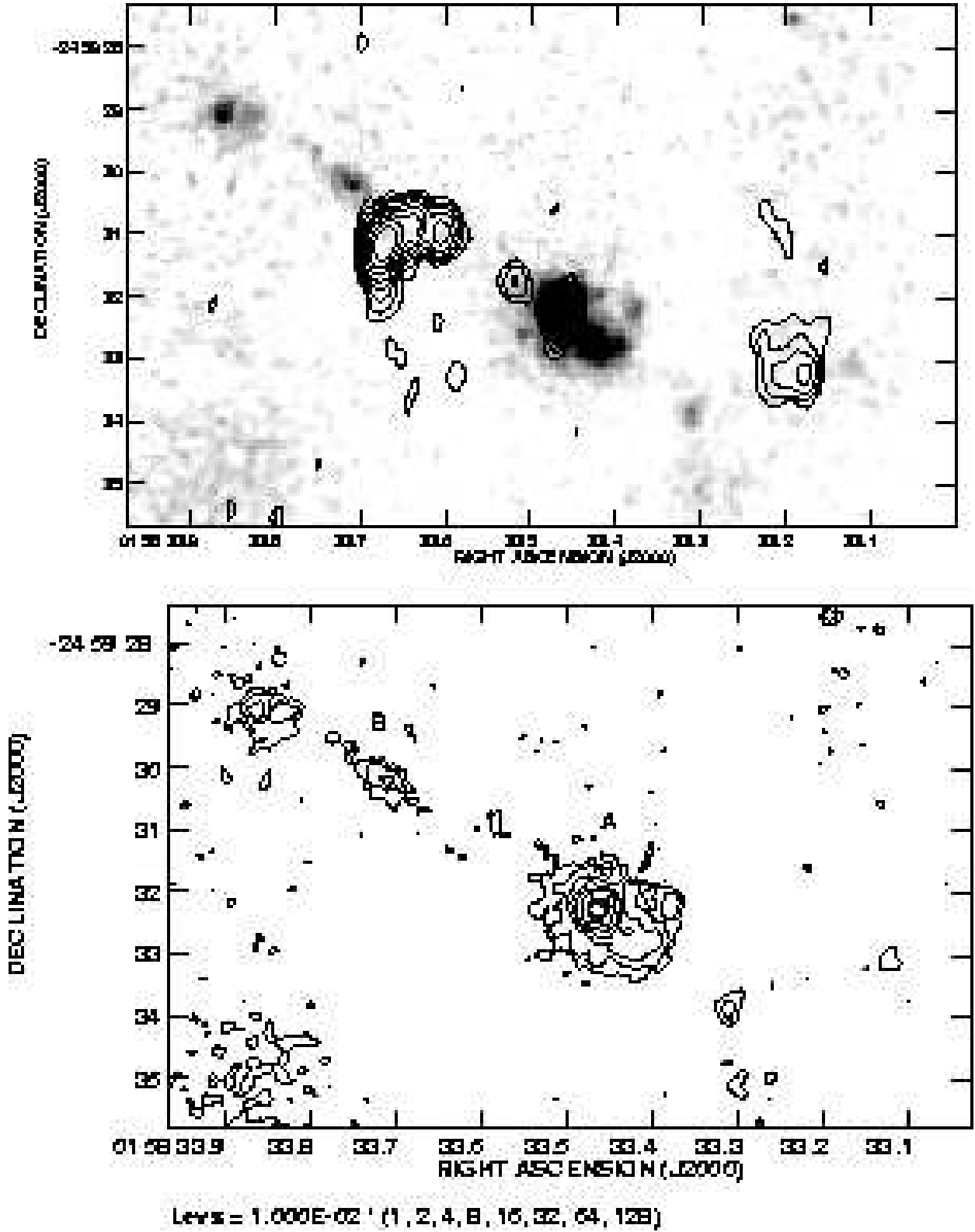}   
\caption[f3.ps]{{\it Top panel}: 
A grey scale representation of the near infrared continuum
emission of the radio galaxy MRC 0156$-$252 at z $=2.09$ with contours
from the VLA 8.2 GHz observations super-imposed. {\it Bottom panel}: 
Contour representation of the continuum emission. 
Contour representation of the same
image.  Contour 
levels of flux density at: 1,2,4,8,16,32,64,128 $\times$ 10$^{-2}$ $\mu$Jy.
{\label{f:0156}}}\end{figure} 
\begin{figure}
\plotone{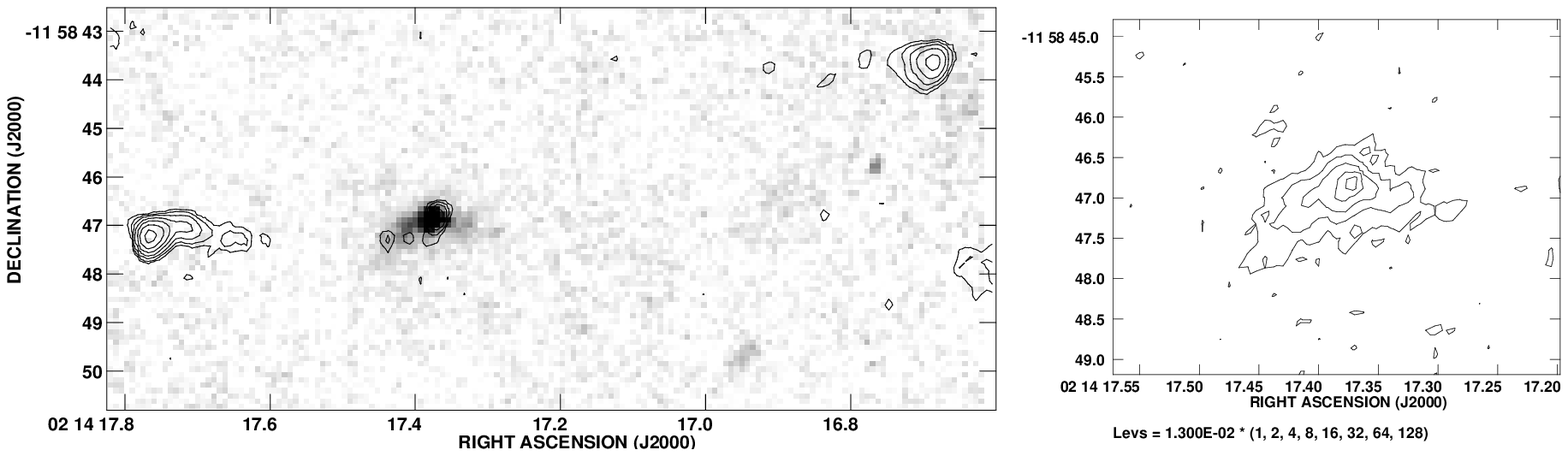}  
\caption[f4.ps]{{\it Left}: 
A grey scale representation of the near infrared continuum
emission of the radio galaxy  USS 0211$-$122 at z $=2.34$ with contours  from
the VLA 8.2 GHz observations super-imposed. 
{\it Right}: Contour representation of 
the continuum emission from the host galaxy.
 Contour 
levels of flux density at: 1,2,4,8,16 $\times$1.3$\cdot$10$^{-2}$ $\mu$Jy.
 {\label{f:0211}}}\end{figure} 
\begin{figure}
\plotone{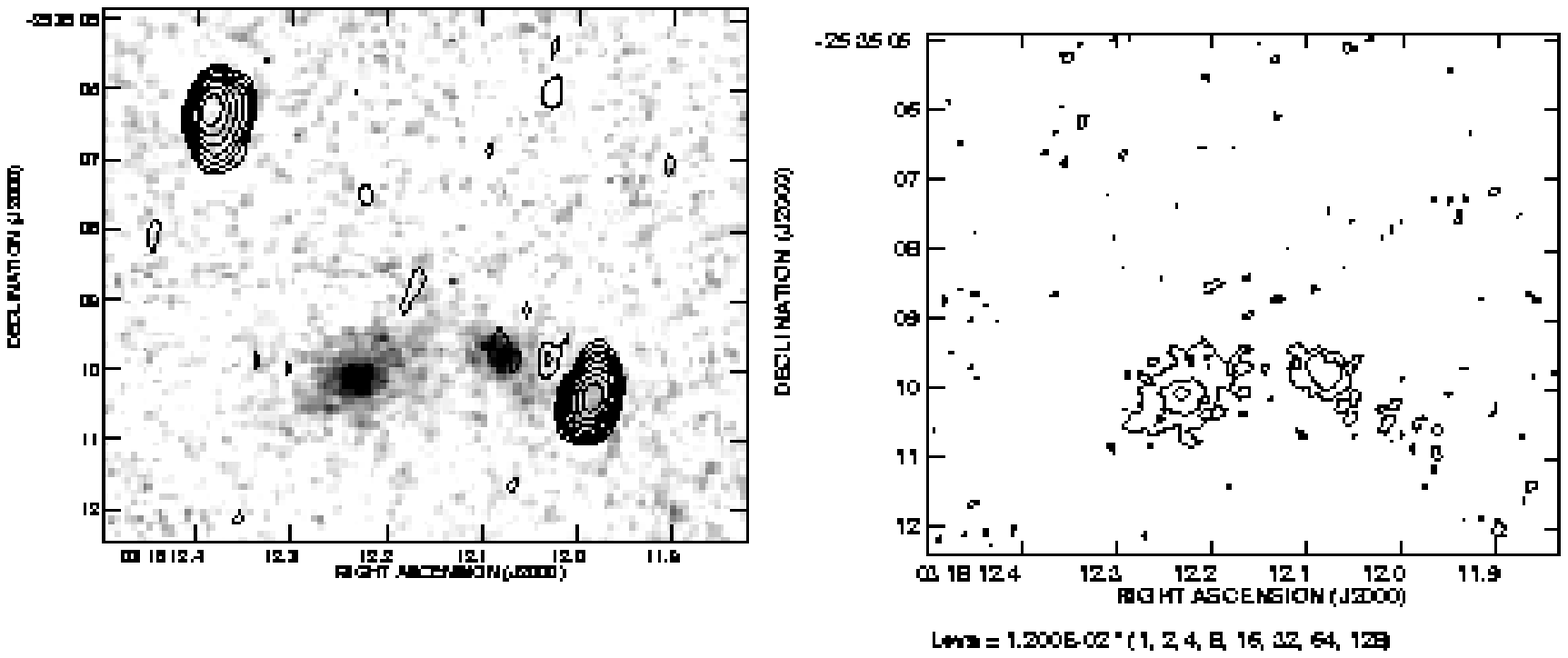}   
\caption[f5.ps]{{\it Left}: A grey scale representation 
of the near infrared continuum
emission of the radio galaxy MRC 0316$-$257 at z $=3.13$ with contours
from the VLA 8.2 GHz observations super-imposed. {\it Right}: 
Contour representation of the continuum emission.
 Contour 
levels of flux density at: 1,2,4 $\times$ 1.2$\cdot$10$^{-2}$ $\mu$Jy.
 {\label{f:0316}}}\end{figure} 

\begin{figure}
\plotone{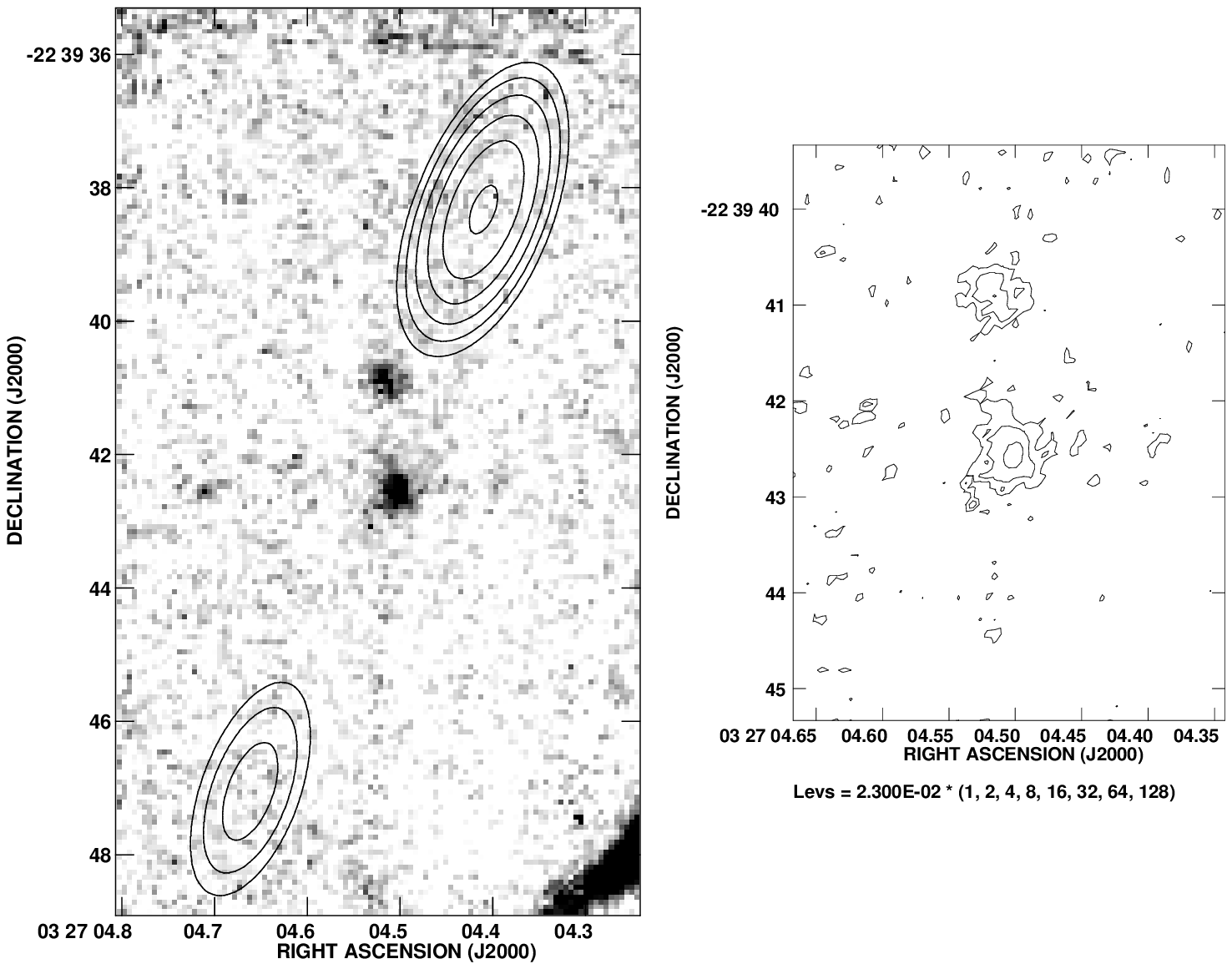} 
\caption[f6.ps]{{\it Left}: A grey scale representation 
of the near infrared continuum
emission of the radio galaxy  MRC 0324$-$228 at z $=1.89$  with contours
from the VLA 4.5 GHz observations super-imposed. {\it Right}: Contour representation of 
the continuum emission from the two possible identifications of the
 host galaxy. Contour 
levels of flux density at: 1,2,4,8 $\times$ 1.3$\cdot$10$^{-2}$ $\mu$Jy. 
{\label{f:0324}}}\vskip1.2cm
\plotone{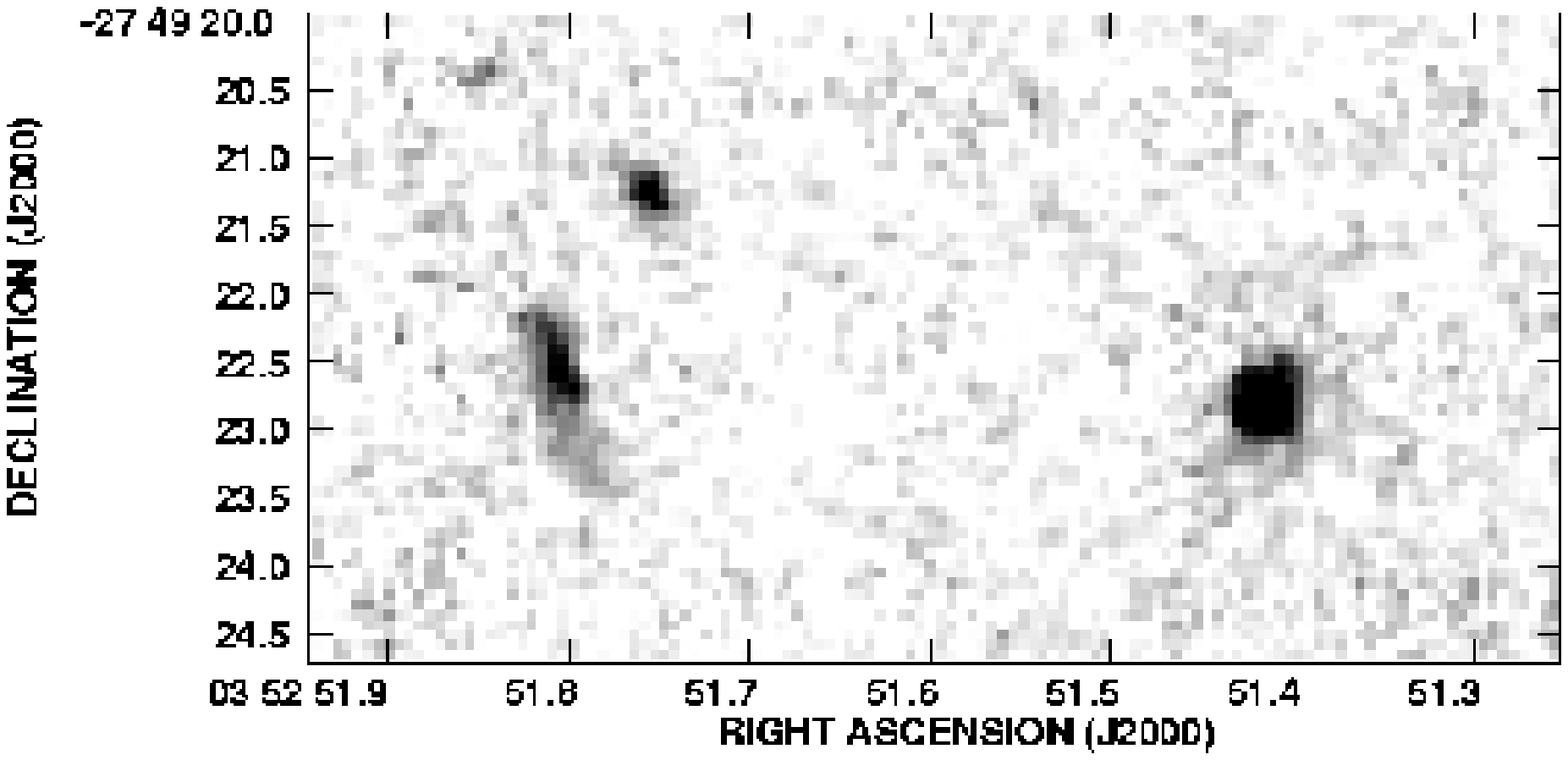} 
\figcaption[f7.ps]
{A grey scale representation of the near infrared continuum
emission of the radio galaxy MRC 0350$-$279 at z$=1.90$. 
{\label{f:0350}}}\end{figure} 
\begin{figure}
\plotone{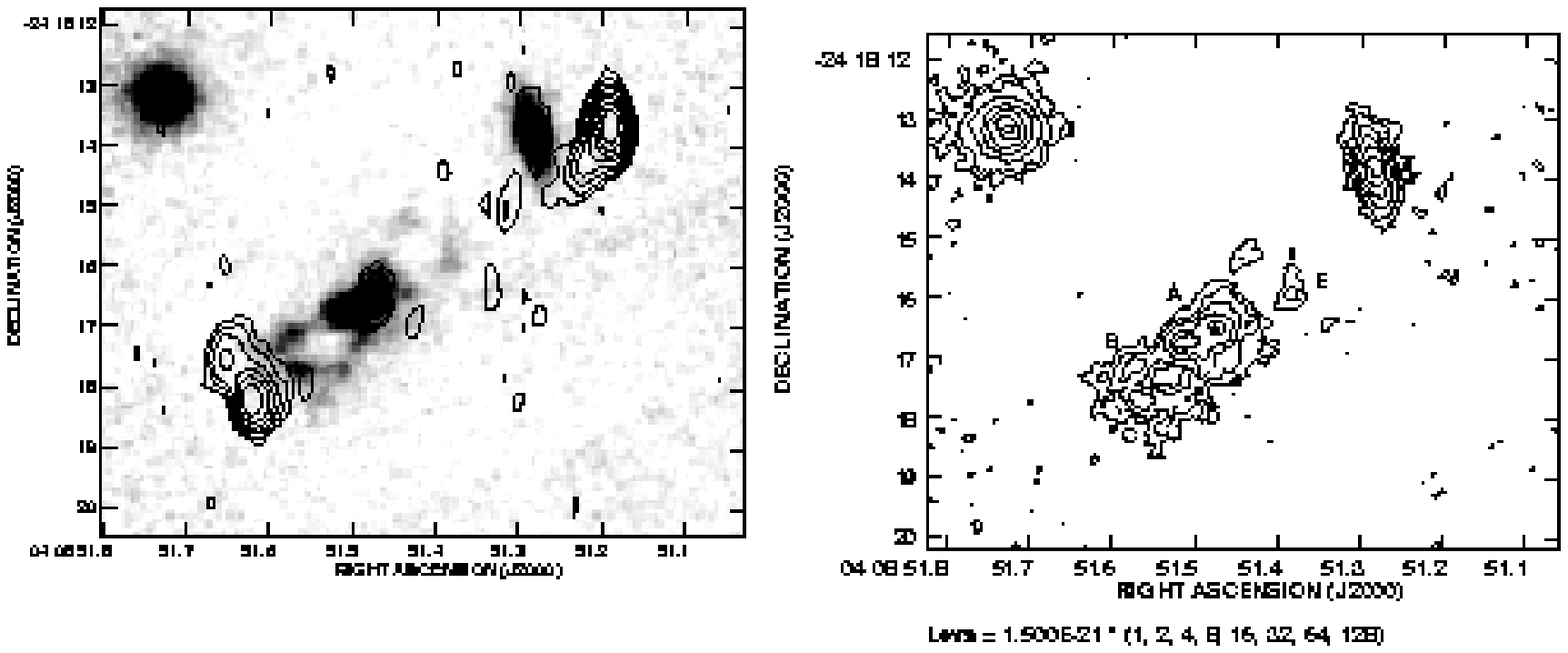} 
\caption[f8.ps]{{\it Left}: A grey scale representation of the near 
infrared continuum
emission of the radio galaxy MRC 0406$-$244 at z $=2.44$  with contours
from the VLA 8.2 GHz observations super-imposed.{\it Right}: 
Contour representation of 
the continuum emission. Contour 
levels of flux density at: 1,2,4,8,16,32,64,128 $\times$
1.5$\cdot$10$^{-21}$erg sec$^{-1}$ cm$^{-2}$ {\label{f:0406}}}
\plotone{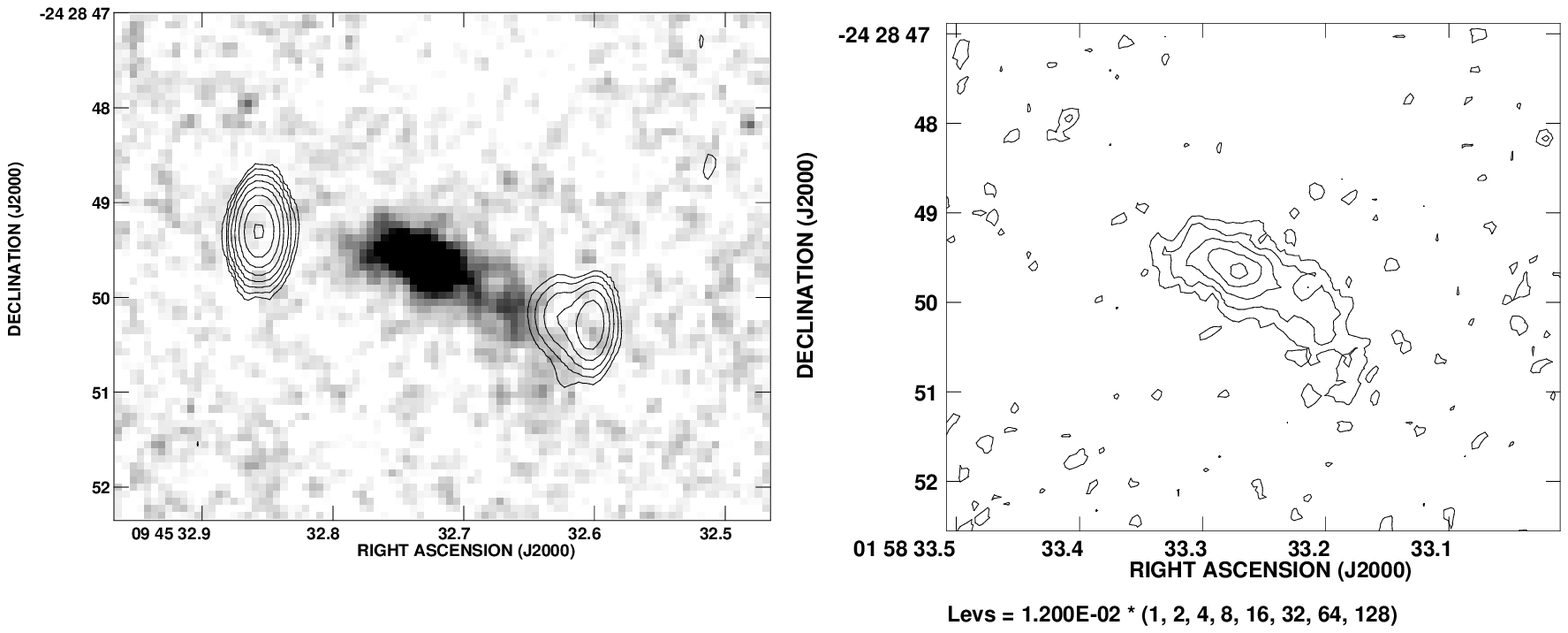} 
\caption[f9.ps]{{\it Left}: A grey scale representation of 
the near infrared continuum
emission of the radio galaxy MRC 0943$-$242 at z $=2.93$ with contours
from the VLA 8.2 GHz
observations super-imposed. {\it Right}: Contour representation of 
the continuum emission. Contour 
levels of flux density at: 1,2,4,8,16 $\times$ 1.2$\cdot$10$^{-2}$ $\mu$Jy.{\label{f:0943}}} \end{figure} 
\begin{figure}
\plotone{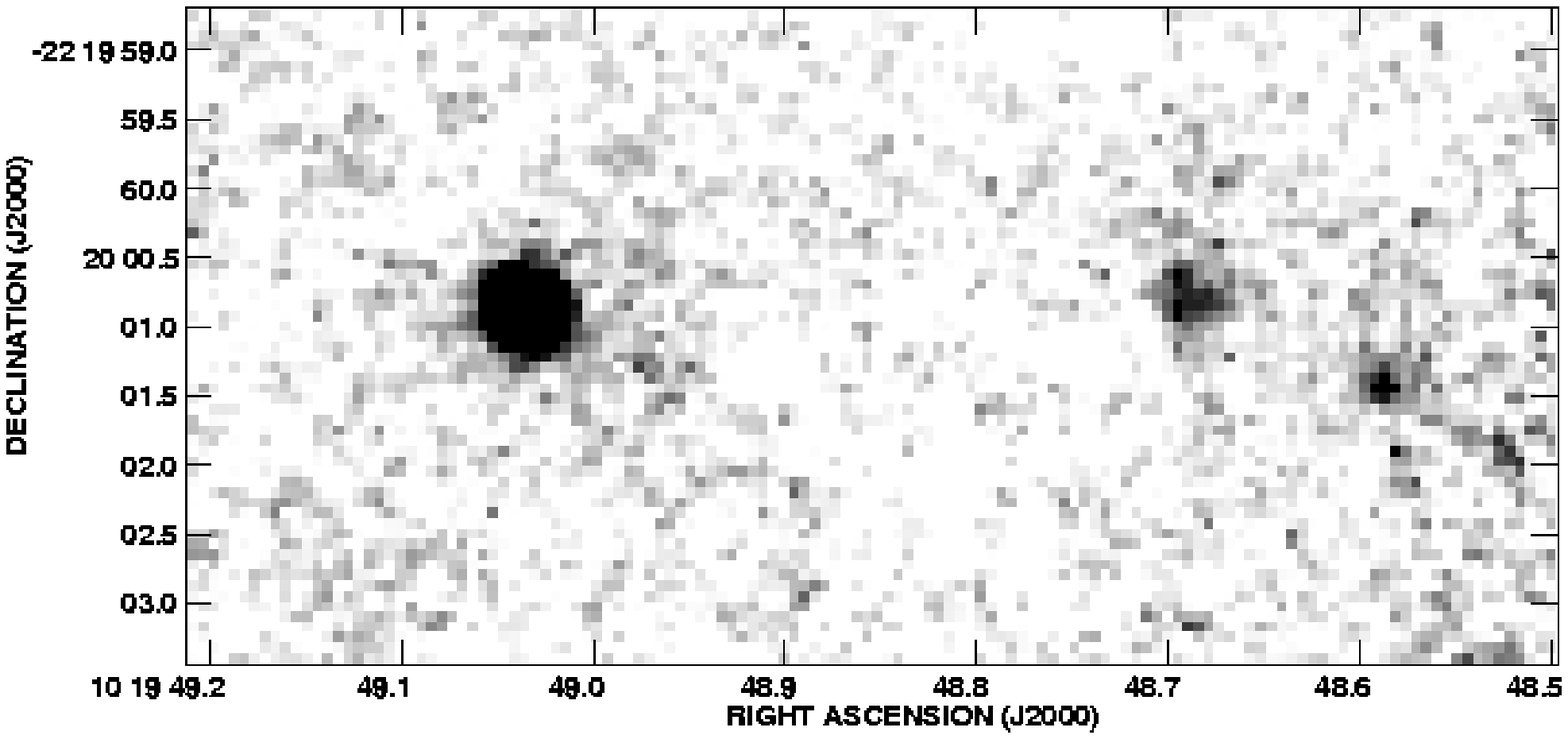} 
\caption[f10.ps]{Grey scale representation of the near infrared continuum
emission of the radio galaxy MRC 1017$-$220 at z $=1.77$.
The host of the radio source is the unresolved object to the east 
{\label{f:1017}}}\vskip1.3cm
\plotone{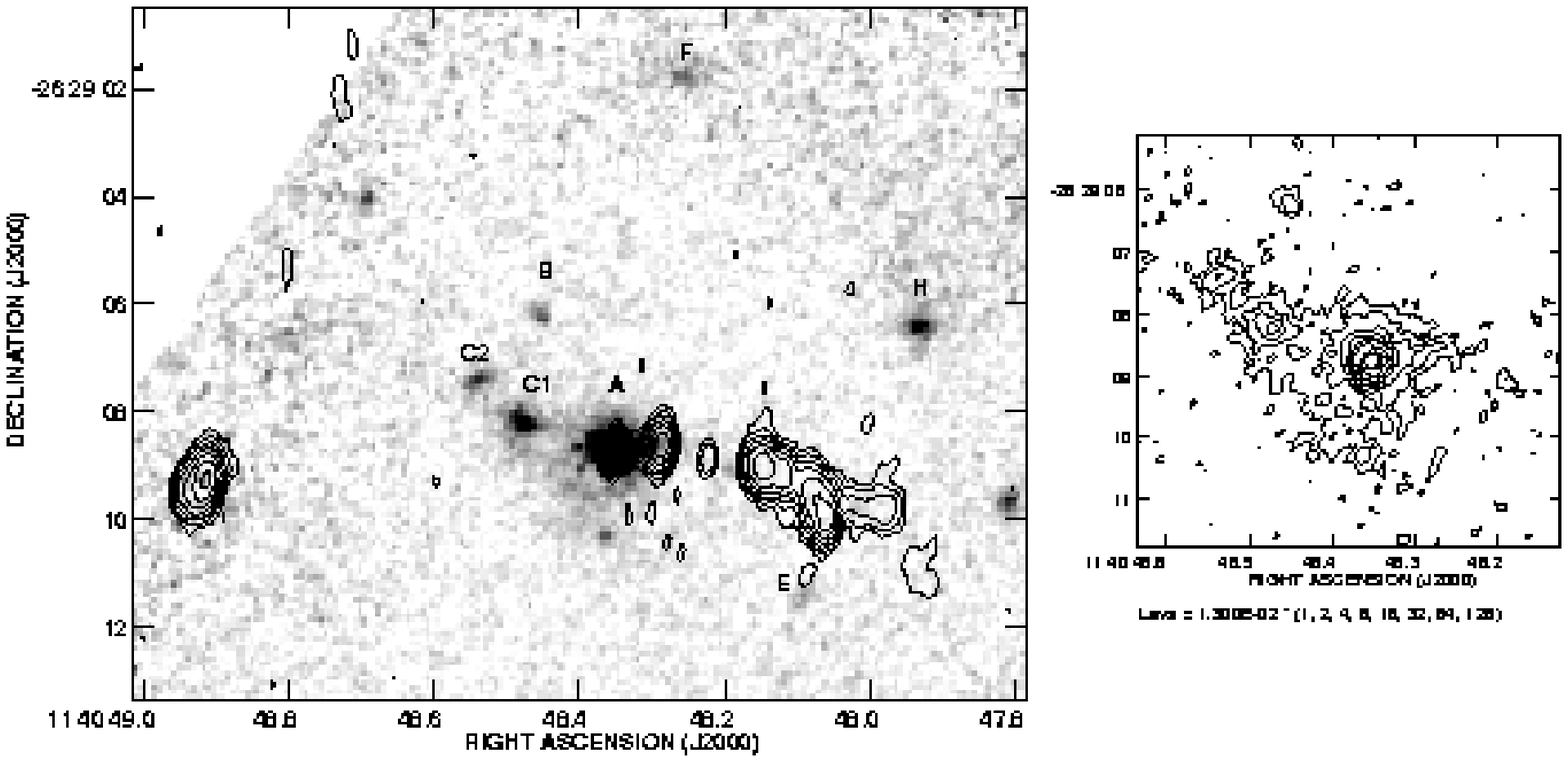} 
\caption[f11.ps]{{\it Left}: A grey scale representation of the near 
infrared continuum
emission of the radio galaxy MRC 1138$-$242 at z $=2.16$ and its close
companions, 
with contours  from the VLA 8.2 GHz
observations super-imposed. {\it Right}: Contour representation of 
the continuum emission from the central region. Contour 
levels of flux density at: 1,2,4,8,16,32,64,128 $\times$ 1.3$\cdot$10$^{-2}$ $\mu$Jy. 
{\label{f:1138}}}\end{figure}\begin{figure}
\plotone{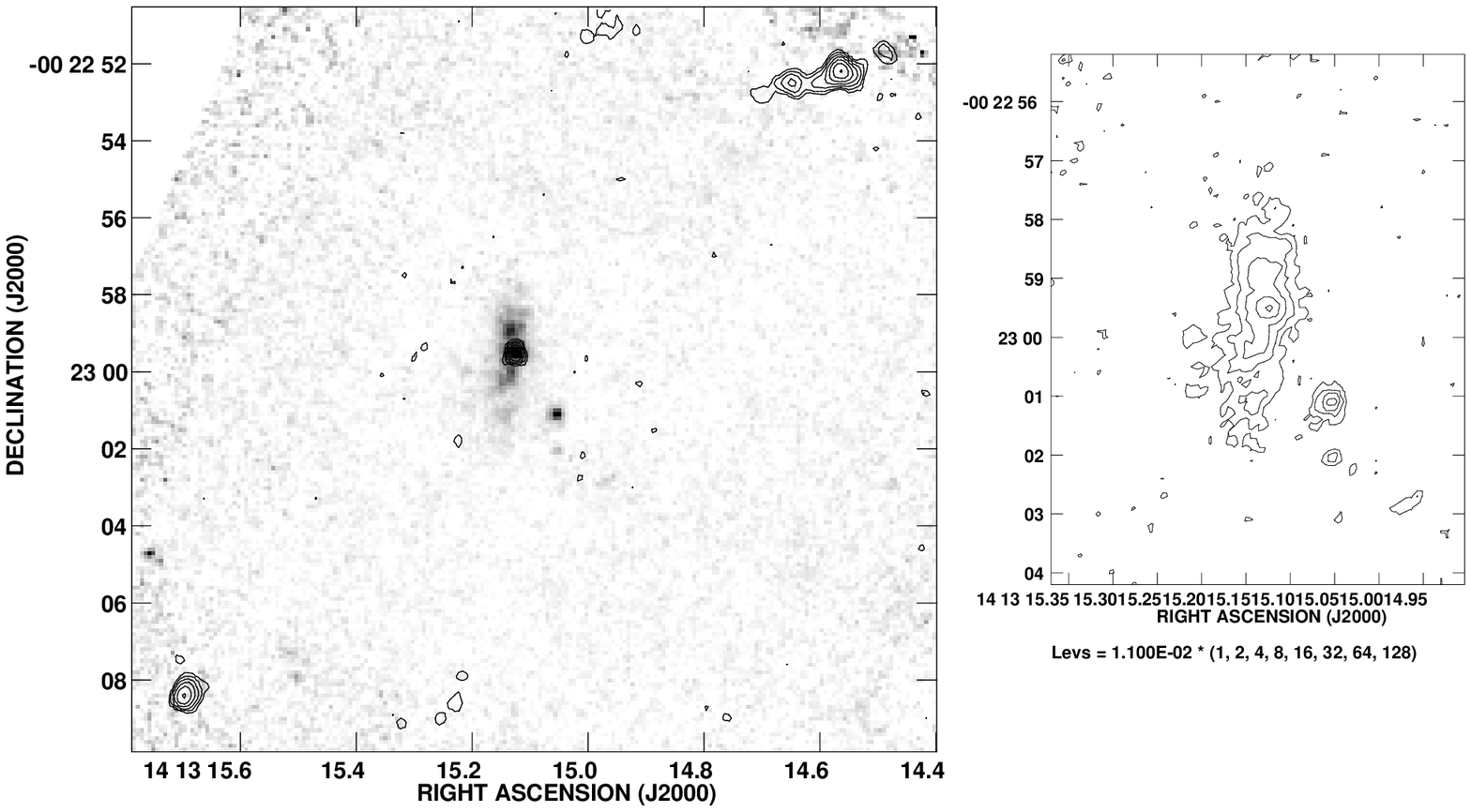} 
\caption[f12.ps]{{\it Left}: Grey
scale representation of the near infrared continuum emission of
the radio galaxy USS 1410$-$001 at z $=2.33$ with contours from the VLA 8.2
GHz observations super-imposed. {\it Right}: Contour representation of 
the continuum emission from the host galaxy. Contour 
levels of flux density at: 1,2,4,8,16 $\times$ 1.1$\cdot$10$^{-2}$ $\mu$Jy.
{\label{f:1410}}}\vskip1.5cm
\plotone{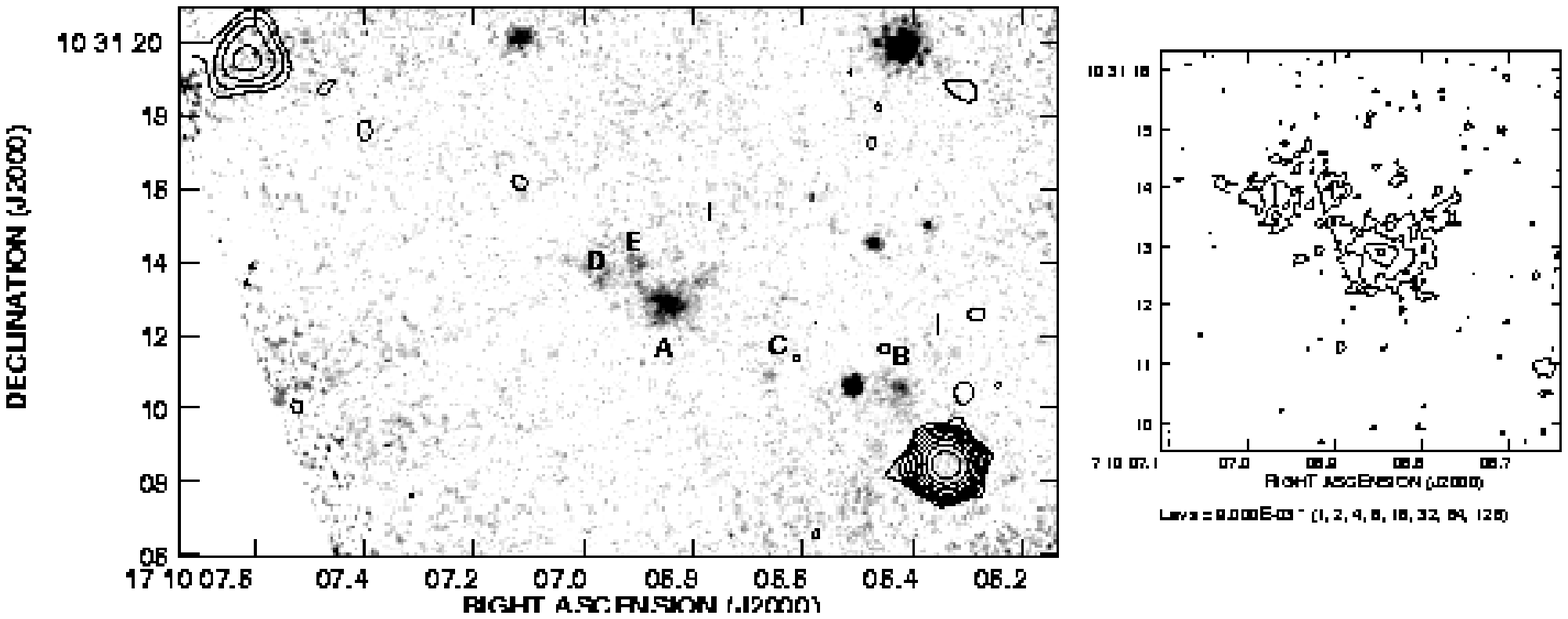} 
\caption[f13.ps]{{\it Top panel}: Grey
scale representation of the near infrared continuum emission of
the radio galaxy USS 1707$+$105 at z $=2.35$ and its close companions,
with contours from
the VLA 8.2 GHz observations super-imposed. {\it Bottom panel}: 
Contour representation of 
the continuum emission from the host galaxy.  Contour 
levels of flux density at: 1,2,4,8,16 $\times$ 0.9$\cdot$10$^{-2}$ $\mu$Jy.
{\label{f:1707}}}\end{figure} \clearpage 
\begin{figure}
\plotone{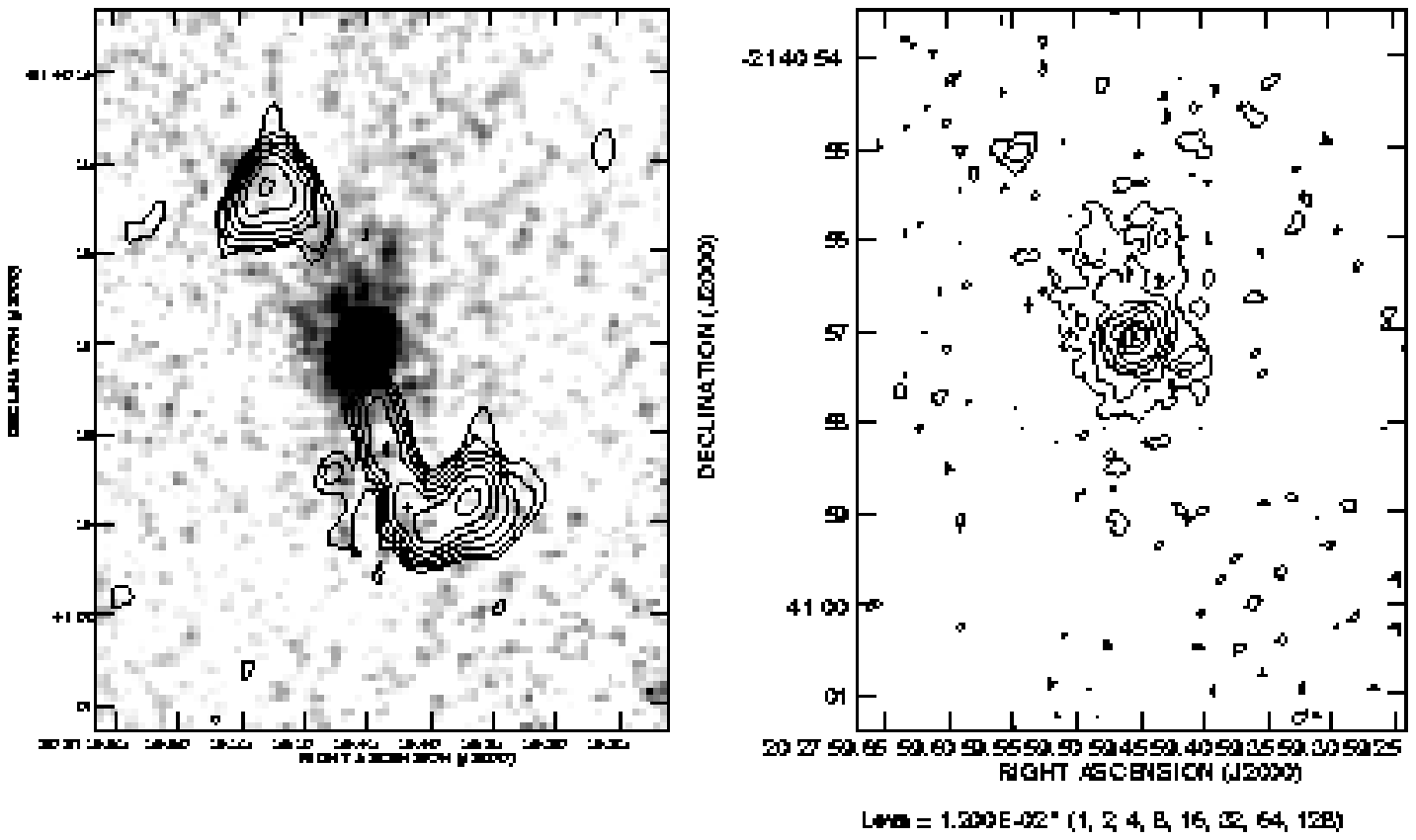} 
\caption[f14.ps]{{\it Left}: Grey scale representation of the near infrared 
continuum
emission of the radio galaxy MRC 2025$-$218 at z $=2.63$ with contours
from the VLA 8.2 GHz
observations super-imposed. {\it Right}: Contour representation of 
the continuum emission. 
Contour 
levels of flux density at: 1,2,4,8,16,32,64,128 $\times$ 1.2$\cdot$10$^{-2}$ $\mu$Jy.
{\label{f:2025}}}
\plotone{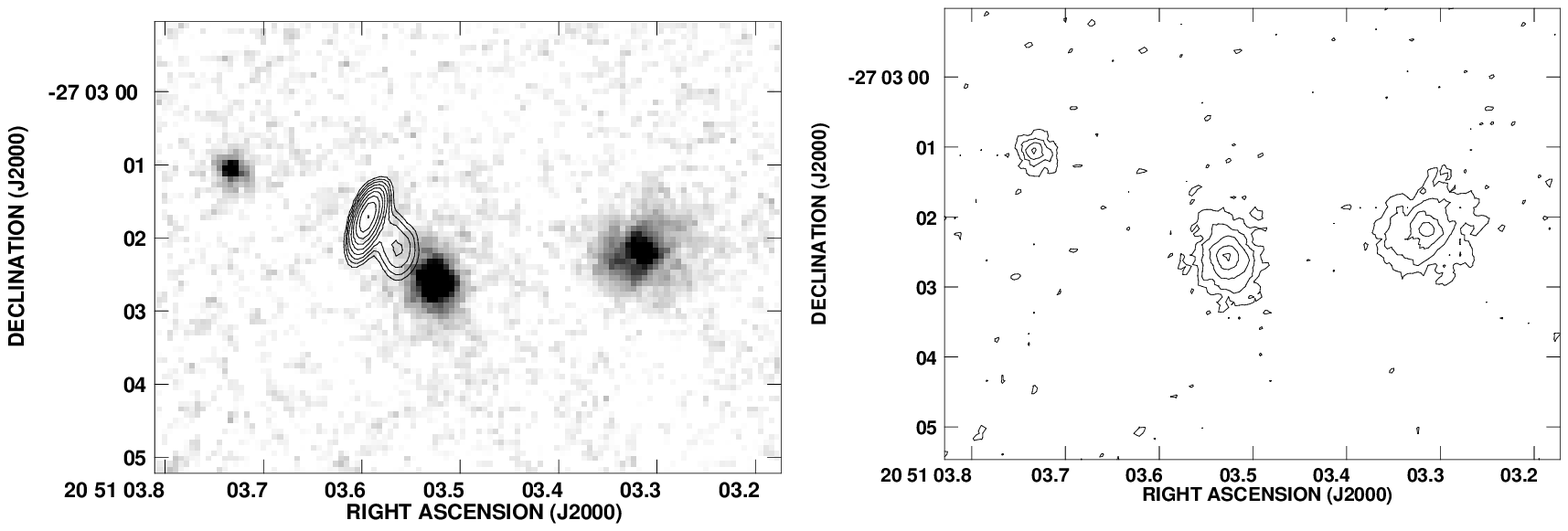} 
\caption[f15.ps]{{\it Left}: Grey scale representation of the near infrared continuum
emission of the radio galaxy MRC 2048$-$272 at z $=2.06$
with contours  from the VLA 8.2 GHz
observations super-imposed. {\it Right}: Contour representation of 
the continuum emission.
Contour 
levels of flux density at: 1,2,4,8,16 $\times$ 1.4$\cdot$10$^{-2}$ $\mu$Jy.
{\label{f:2048}}}\end{figure} 
\clearpage 
\begin{figure}
\plotone{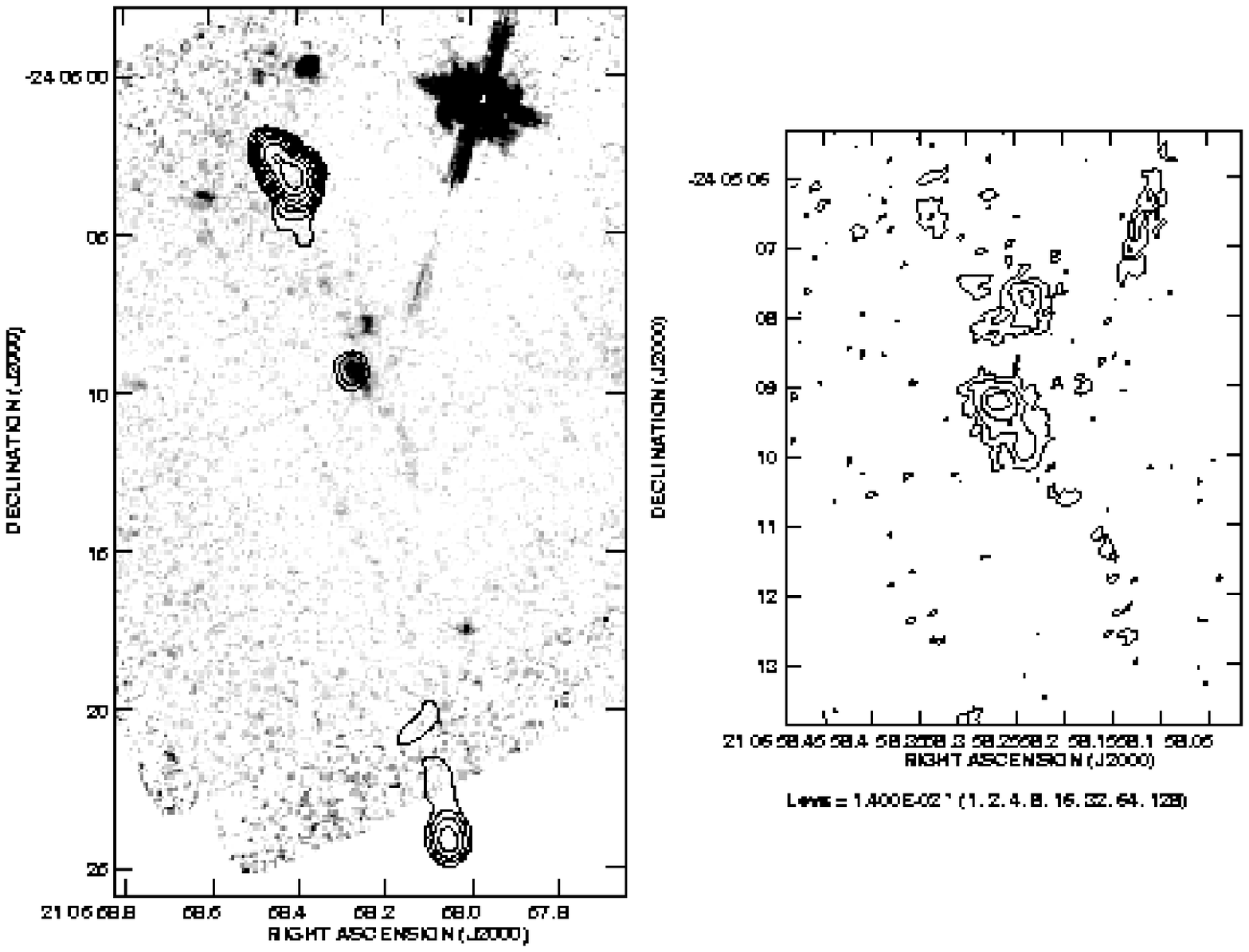} 
\caption[f16.ps]
{{\it Left}: Grey scale representation of the near infrared continuum
emission of the radio galaxy MRC 2104$-$242 at z $=2.49$ with contours  from the VLA 8.2 GHz
observations super-imposed. {\it Right}: Contour representation of 
the continuum emission from the central region. The elongated feature 
north-west of the galaxy core is a spike from the near by star.
Contour 
levels of flux density at: 1,2,4,8,16 $\times$ 1.4$\cdot$10$^{-2}$ $\mu$Jy.
 {\label{f:2104}}}\vskip1.4cm
\plotone{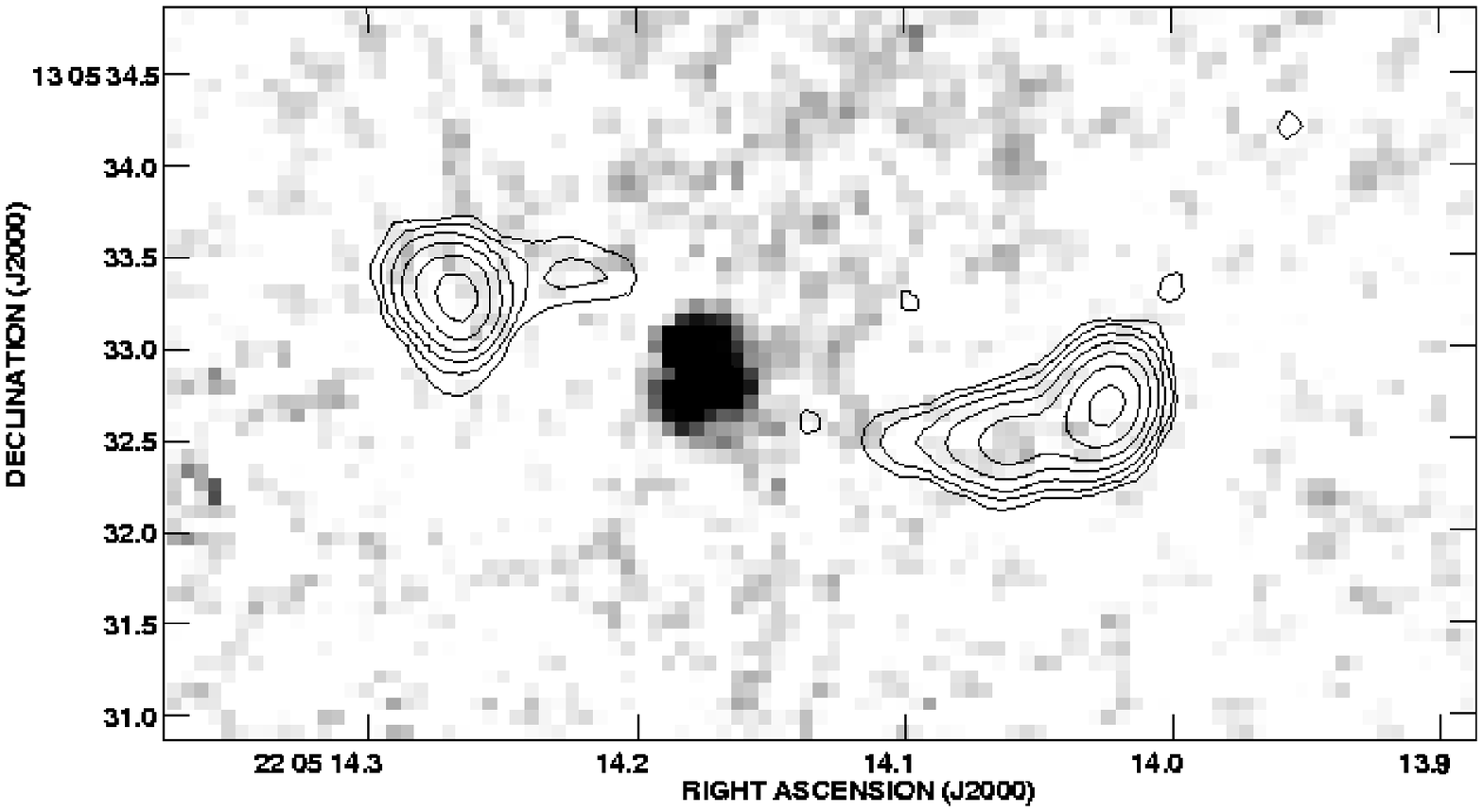} 
\caption[f17.ps]{Grey scale representation of the near infrared continuum
emission of the radio galaxy USS 2202$+$128 at z $=2.70$  with contours  from the VLA 8.2 GHz
observations super-imposed.{\label{f:2202}}}\end{figure}\begin{figure}
\plotone{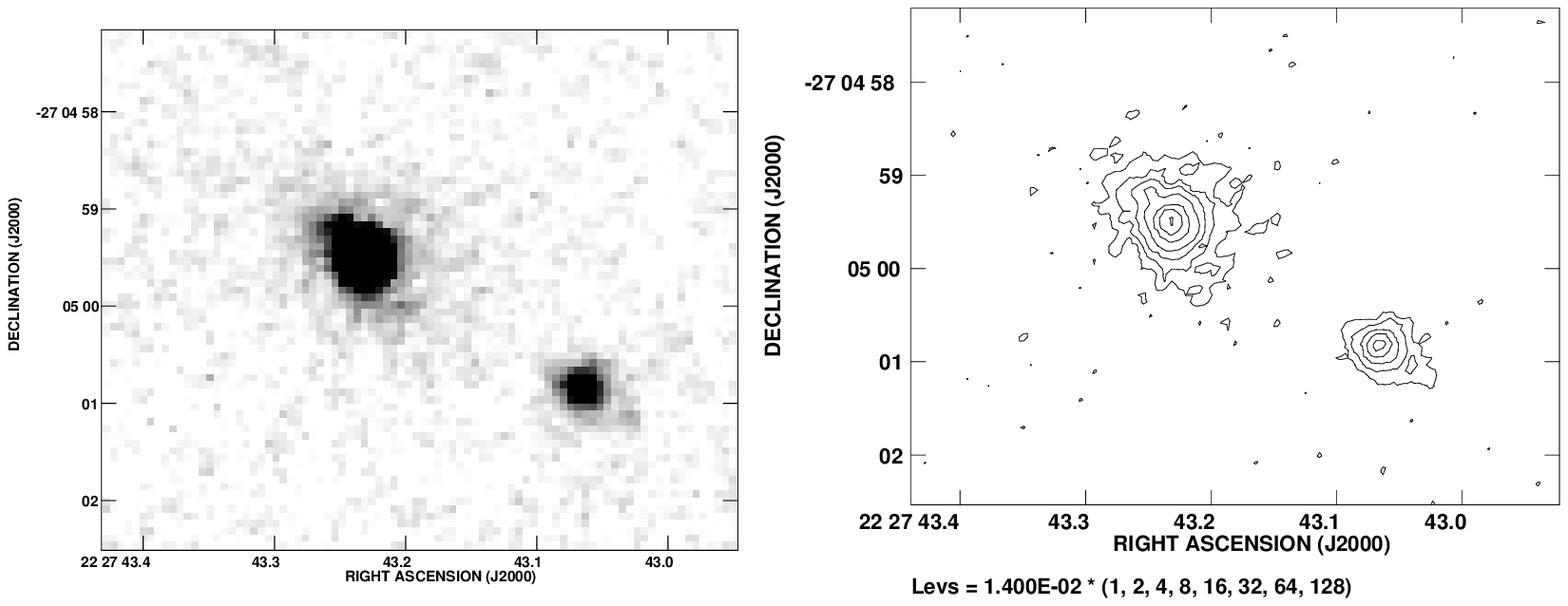} 
\caption[f18.ps]{{\it Left}: Grey scale representation of the near 
infrared continuum
emission of the radio galaxy MRC 2224$-$273 at z $=1.68$. {\it Right}:
Contour representation of 
the continuum emission. 
Contour 
levels of flux density at: 1,2,4,8,16,32,64 $\times$ 1.4$\cdot$10$^{-2}$ $\mu$Jy.
{\label{f:2224}}}
\plotone{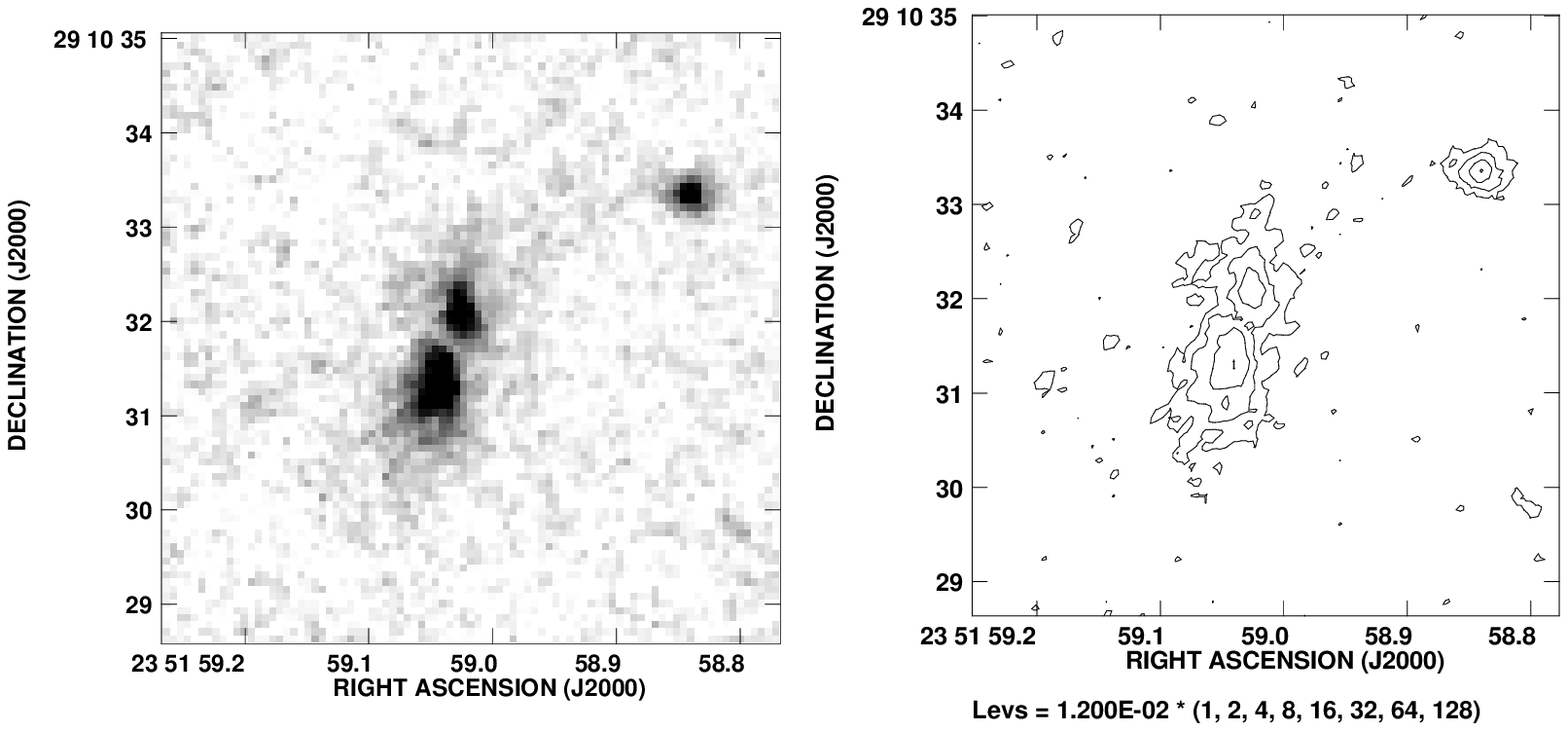} 
\caption[f19.ps]{{\it Left}: Grey scale representation of the near infrared continuum
emission of the radio galaxy USS 2349$+$280 at z $=2.89$. {\it Right}:
Contour representation of 
the continuum emission. Contour 
levels of flux density at: 1,2,4,8 $\times$ 1.2$\cdot$10$^{-2}$ $\mu$Jy. {\label{f:2349}}}\end{figure} 
\begin{figure}
\plotone{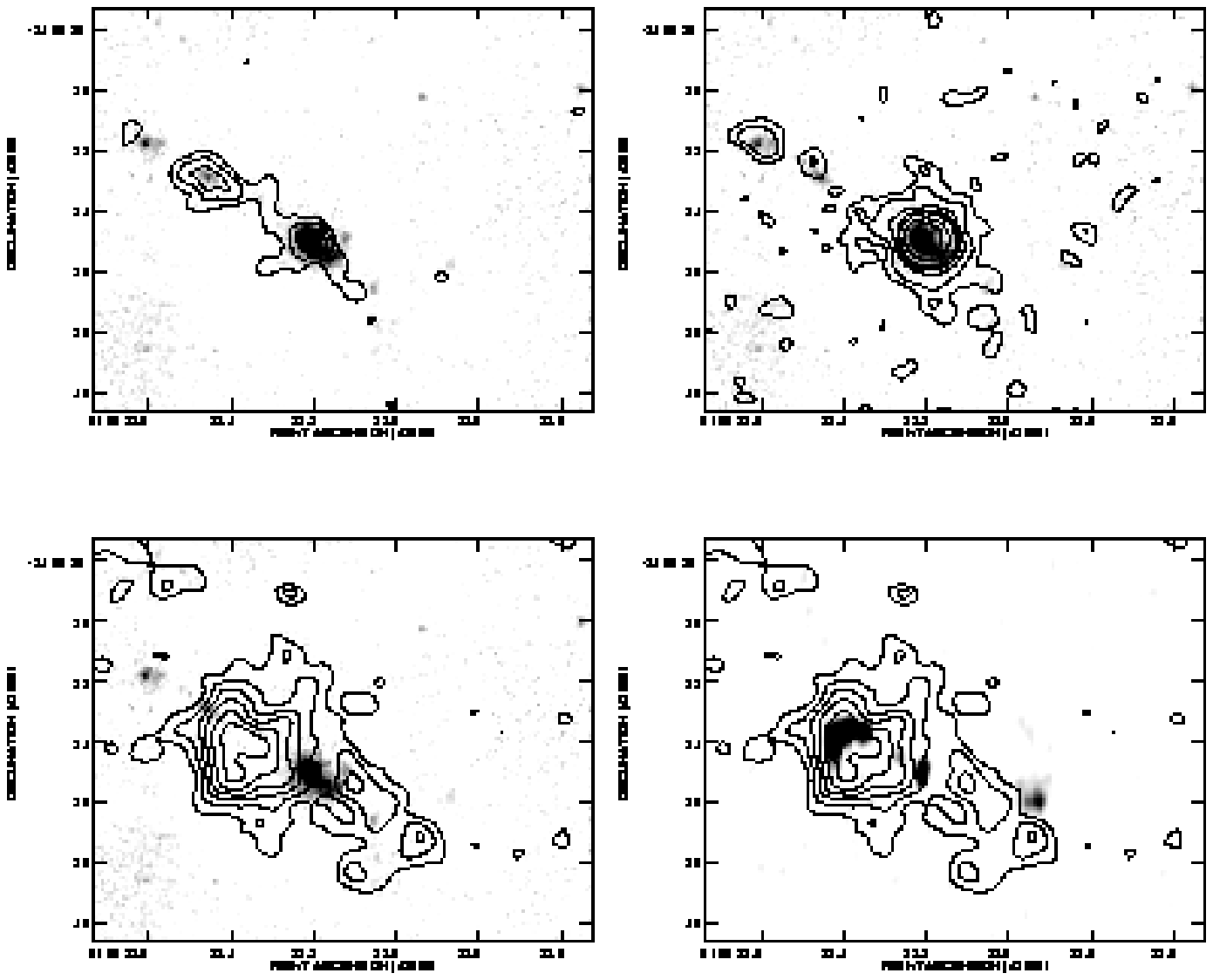} 
\caption[f20.ps]{The NICMOS image of MRC 0156$-$252 in grey scale
overlayed with contours representing the I-band emission (top
left), K-band emission (top right) and narrow band Ly$\alpha$
emission (bottom left). The bottom right panel shows the 8.2 GHz
radio emission in grey scale overlayed with contour of the narrow
band Ly$\alpha$ emission}\label{aa}\end{figure} 
\begin{figure}
\plotone{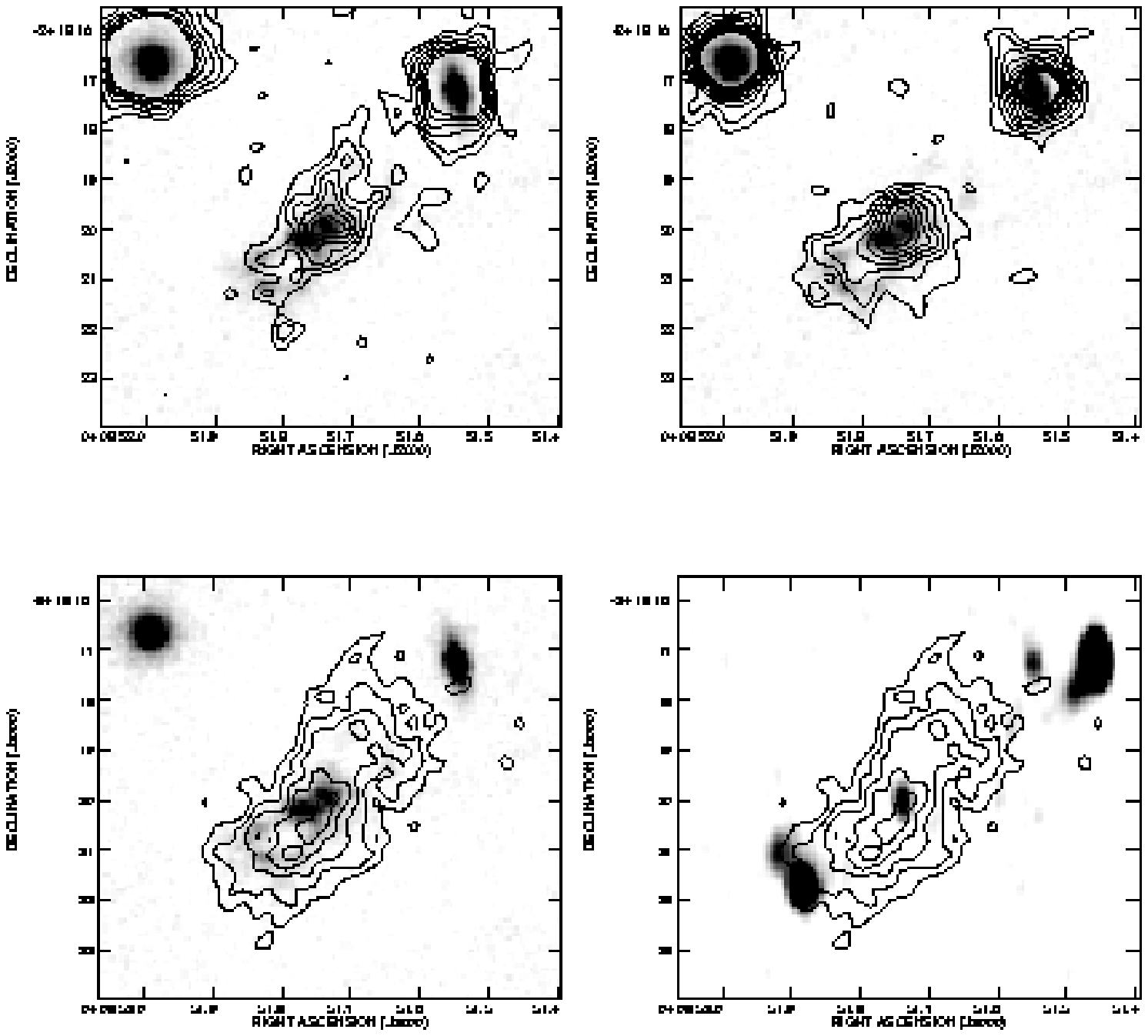} 
\caption[f21.ps]{The NICMOS image of MRC 0406$-$244 in grey scale
overlayed with contours representing the I-band emission (top
left), K-band emission (top right) and narrow band Ly$\alpha$
emission (bottom left).The bottom right panel shows the 8.2 GHz
radio emission in grey scale overlayed with contour of the narrow
band Ly$\alpha$ emission.}\end{figure} 
\begin{figure} 
\plotone{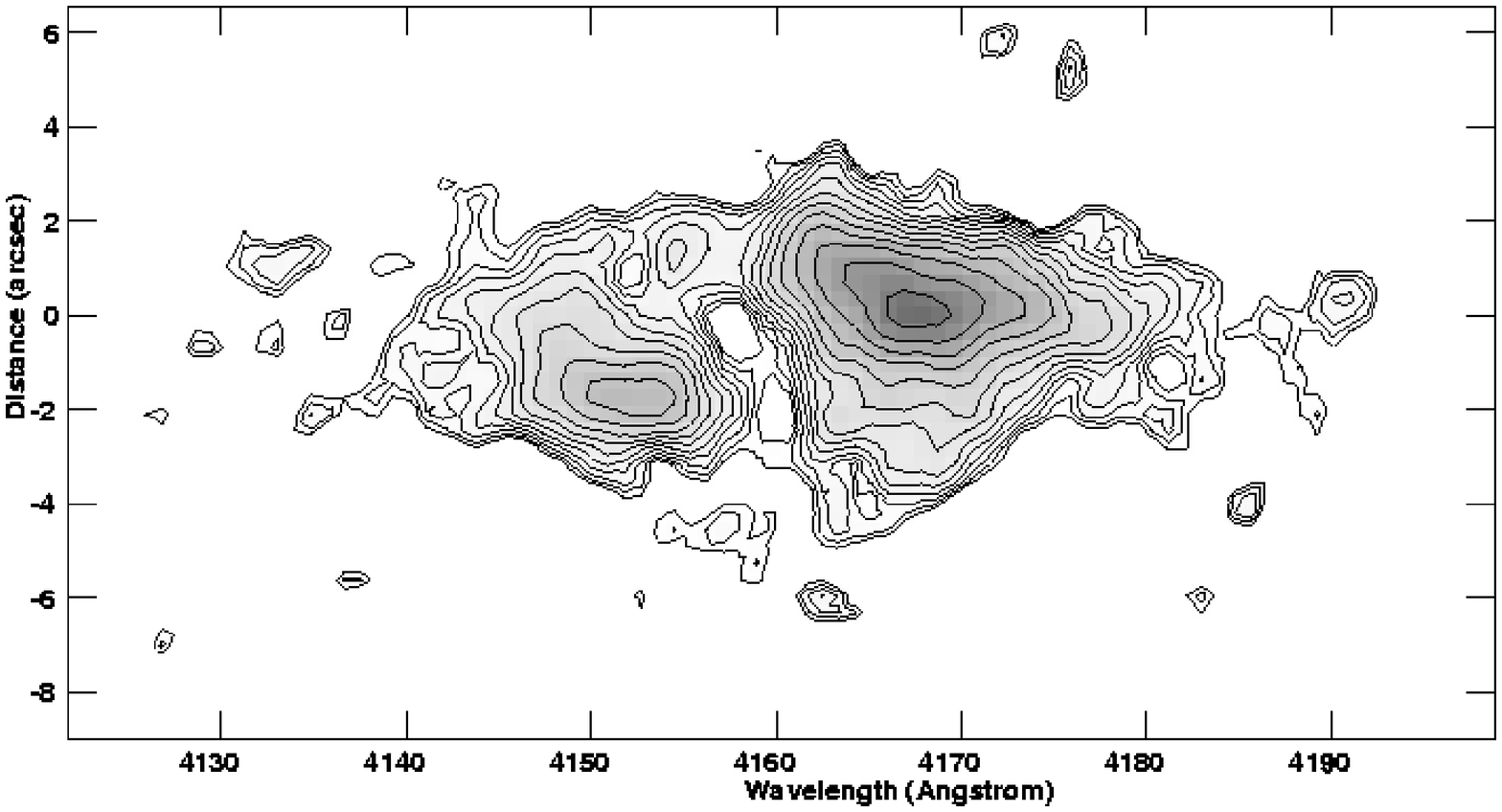} 
\figcaption[f22.ps]{A
high resolution spectrum of the Ly$\alpha$ emission line from
 MRC 0406$-$244, taken with the NTT, having a resolution of 2.8 \AA }\label{bb1}\end{figure} \clearpage
\begin{figure}
\plotone{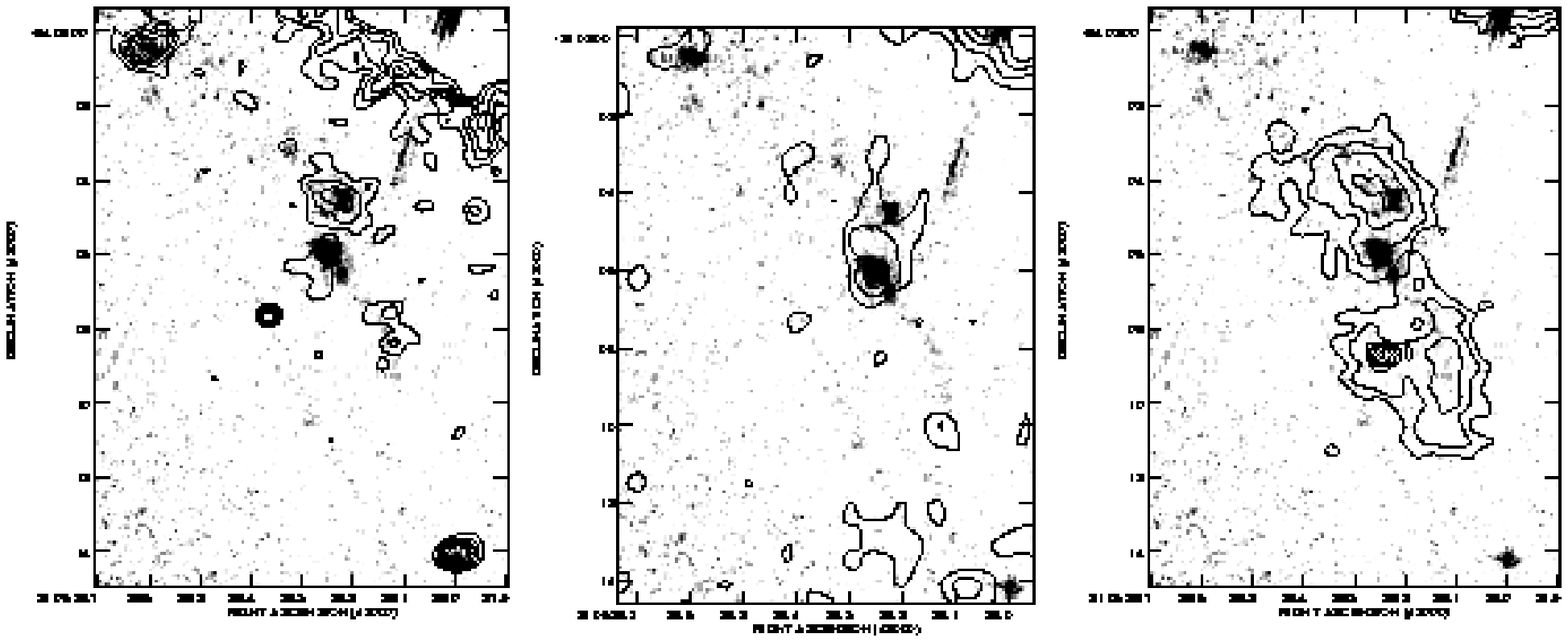} 
\caption[f23.ps]
{The NICMOS image of MRC 2104$-$242 in grey scale overlayed with
contours of the R-band emission (left panel) K-band emission
(central panel), and narrow band Ly$\alpha$ emission (right
panel). The elongated feature on the top right of the images is a
spike from a nearby bright star. }\label{cc} 
\vskip1cm
\plotone{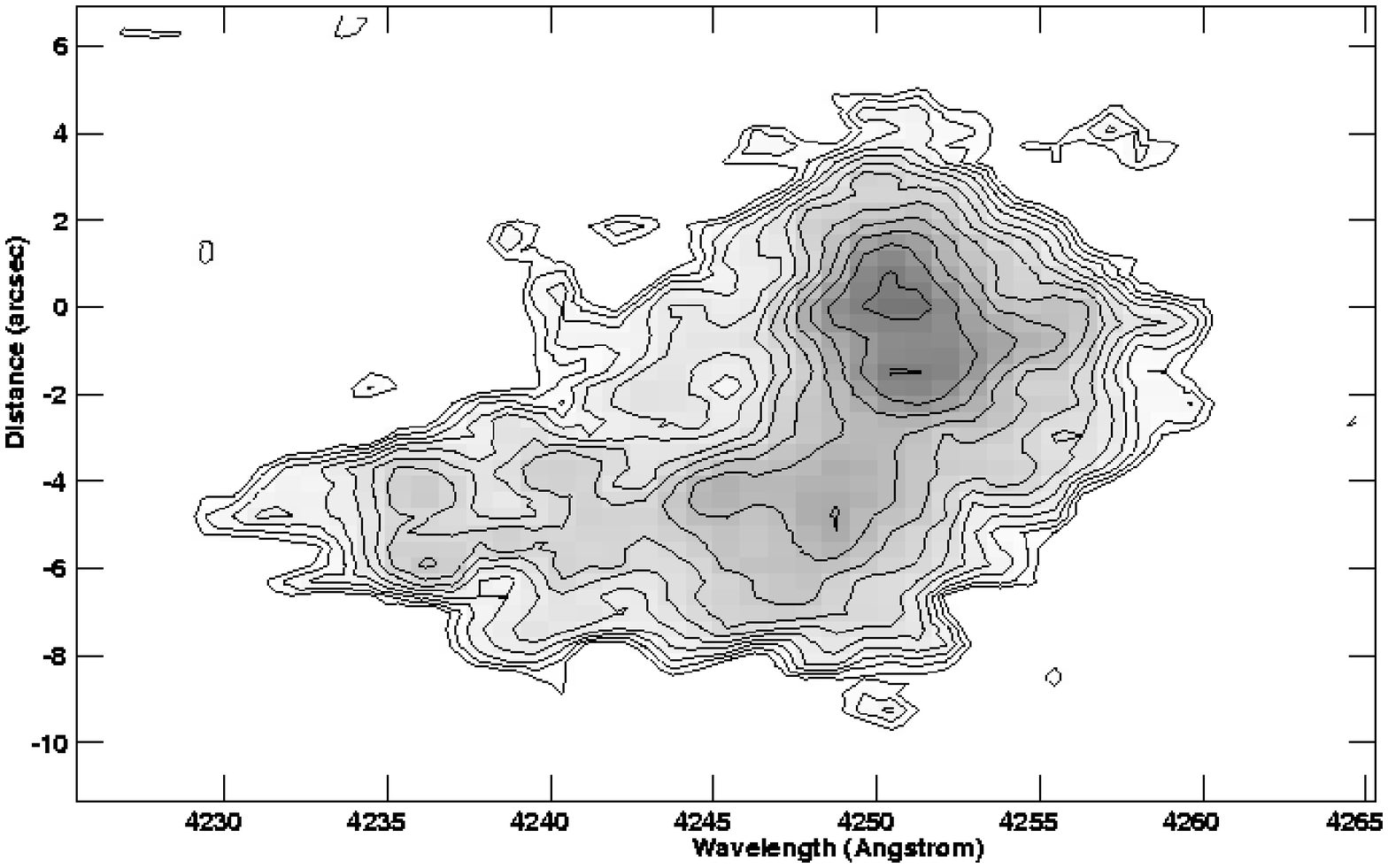} 
\caption[f24.ps]{A
high resolution spectrum of the Ly$\alpha$ emission line from
MRC 2104$-$242, taken with the NTT, having a resolution of 2.8 \AA
}\label{cc1}\end{figure} 
\begin{figure}
\plotone{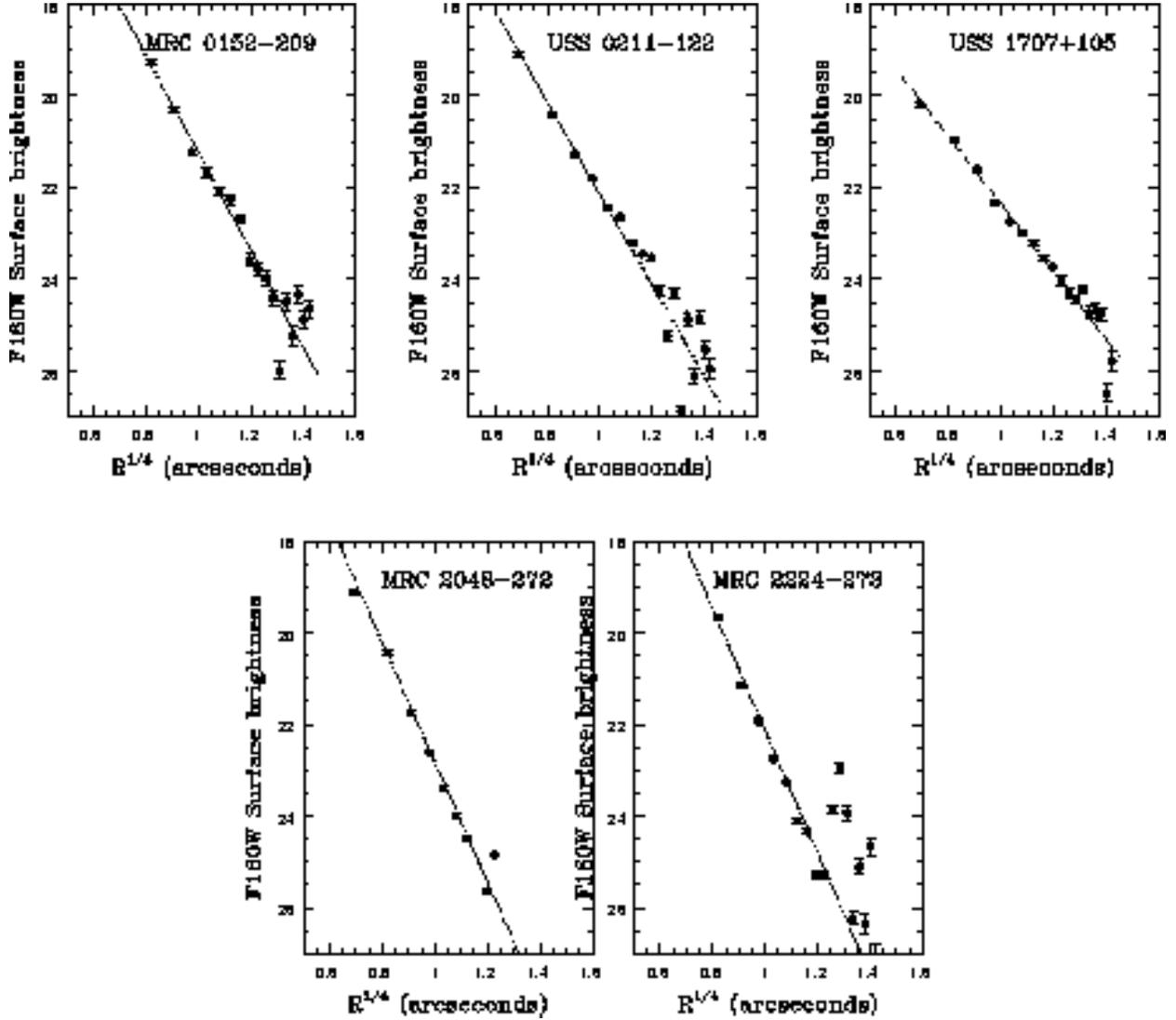} 
\figcaption[f25.ps]{Surface brightness profiles for the 5 radio galaxies
which are well represented by a de Vaucouleurs law: the filled circles are
the data with thier relative error bars, while the dotted lines are the
best fit laws. }\label{f:fit}\end{figure}
\begin{figure}
\plotone{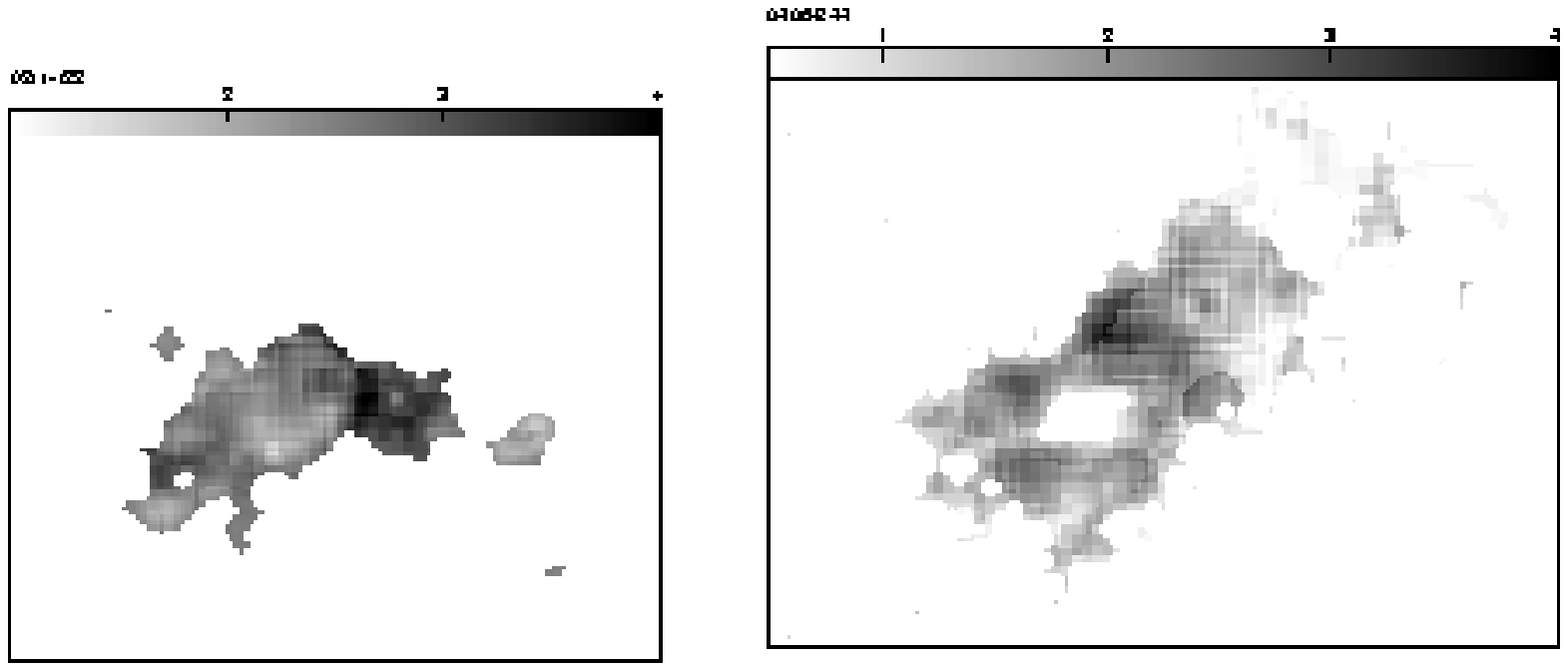} 
\caption[f26.ps]{The
optical to infrared spectral index distribution for the radio
galaxies USS 0211$-$122 (left) and MRC 0406$-$244 (right). The bar on the top
indicates the scale in each case }
\plotone{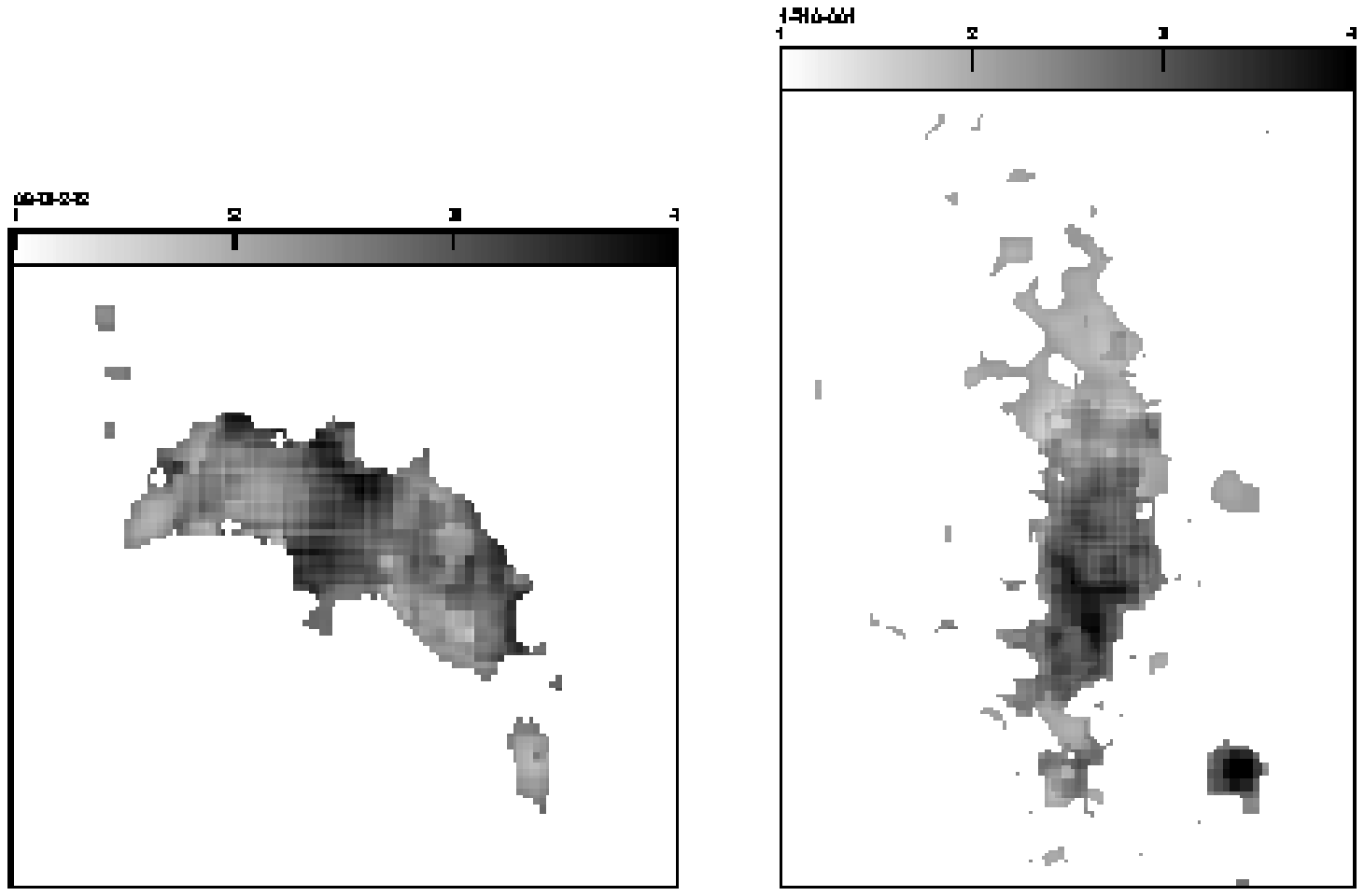} 
\caption[f27.ps]{The
optical to infrared spectral index distribution for the radio
galaxies MRC 0943$-$242 (left) USS 1410$-$001 (right). The bar on the top
indicates the scale in each case }\end{figure}
\begin{figure}
\plotone{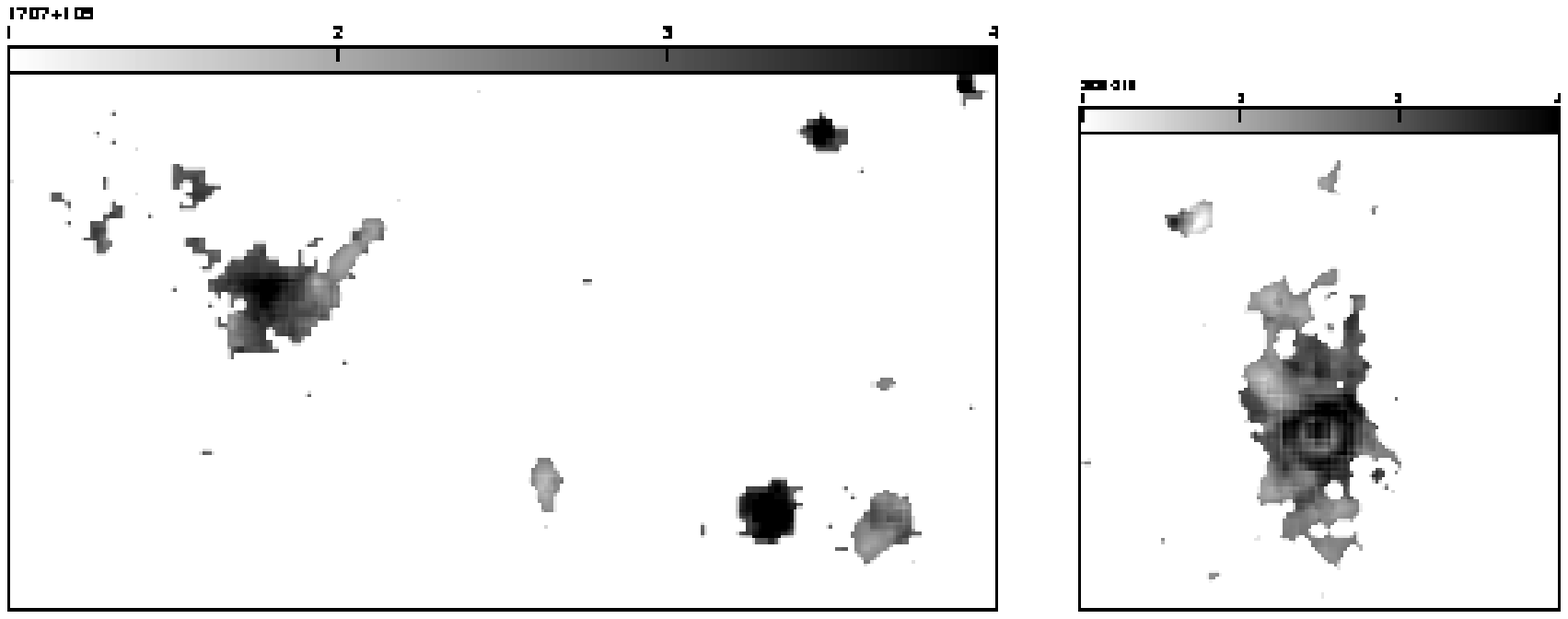} 
\caption[f28.ps]{The optical to infrared spectral index distribution for
the radio galaxy USS 1707+105 (left) and MRC 2025$-$218 (right). The bar on the top indicates the scale
in each case }
\plotone{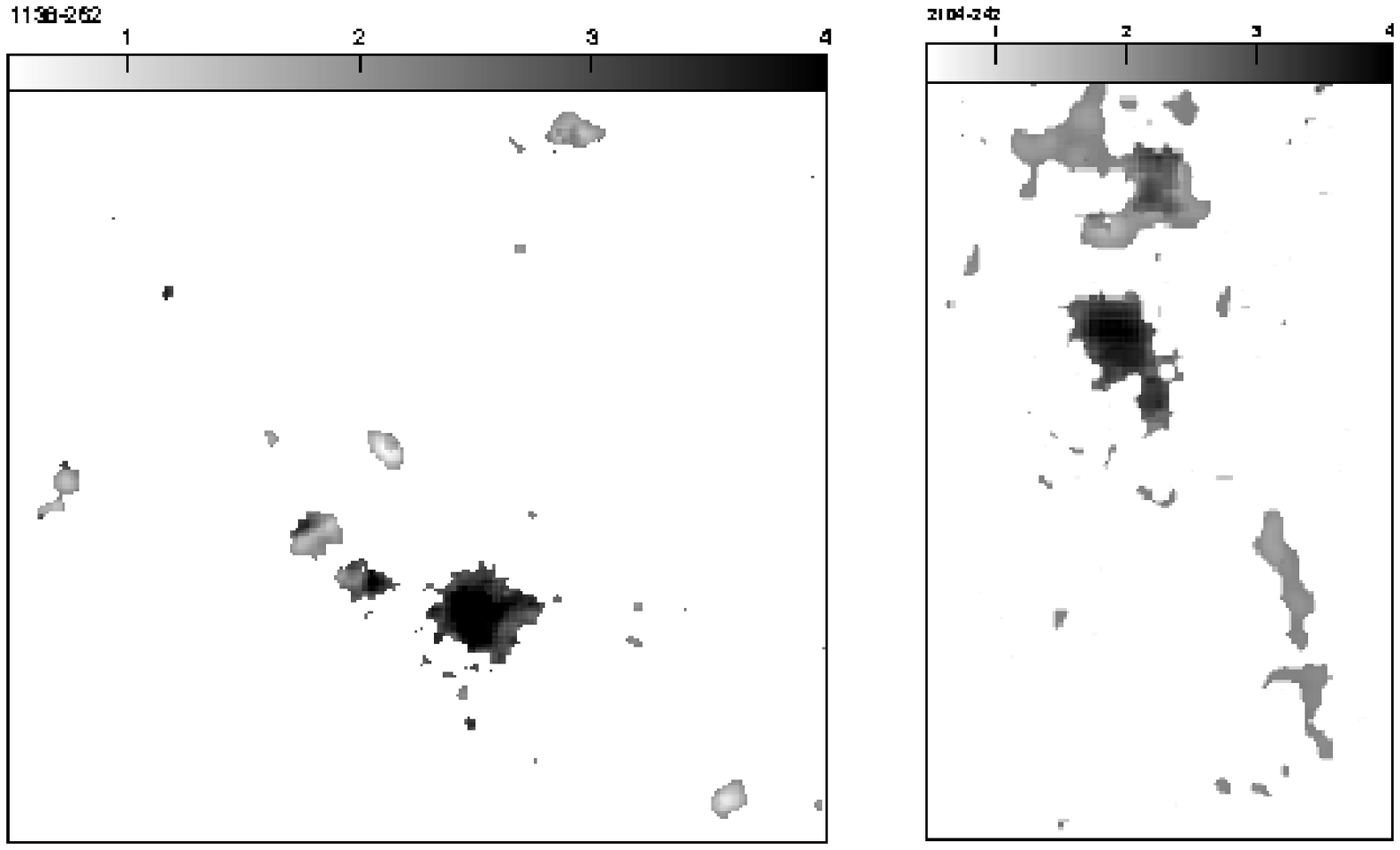}
\caption[f29.ps]{The
optical to infrared spectral index distribution for the radio
galaxies MRC 1138$-$262(left) and MRC 2025$-$218 (right). The bar on the top
indicates the scale in each case. In these galaxies there is a
strong contribution from a nuclear point source which influence
the spectral indexes in the center. }\end{figure}
\begin{figure}
\plotone{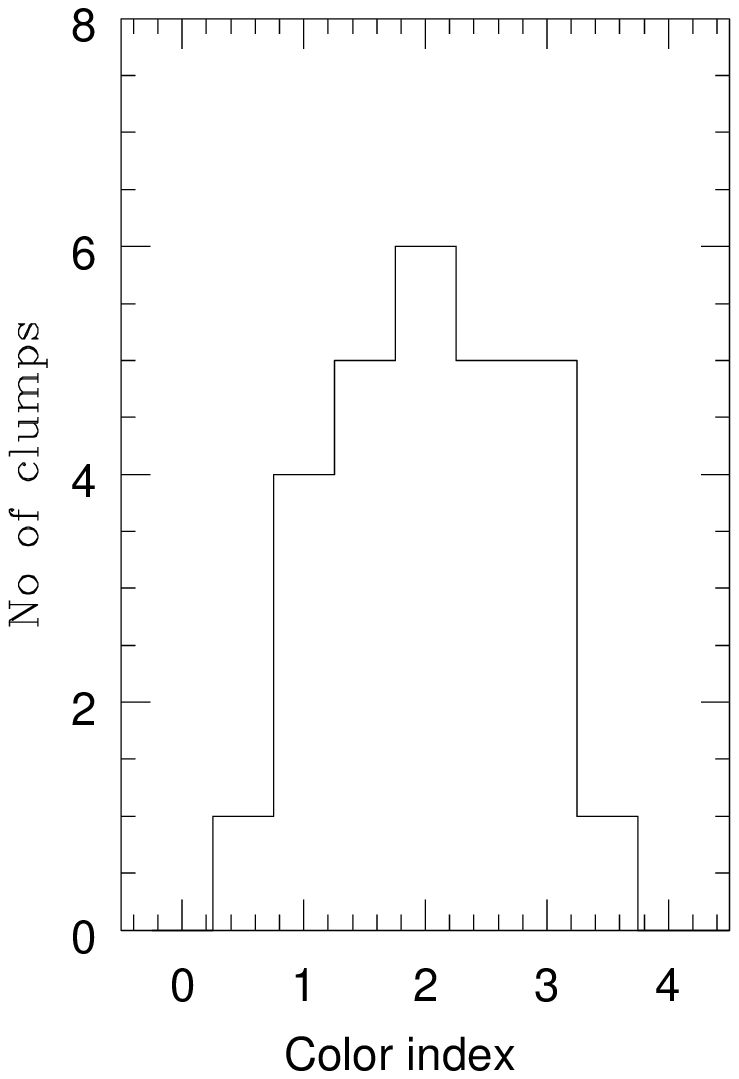}
\caption[f30.ps]{Distribution of the optical to infrared
spectral indices of the components of the 4 clumpy radio galaxies,
MRC 0406$-$244, MRC 1138$-$262, USS 1707$+$105 and MRC 2104$-$242.}\label{isto}

\plotone{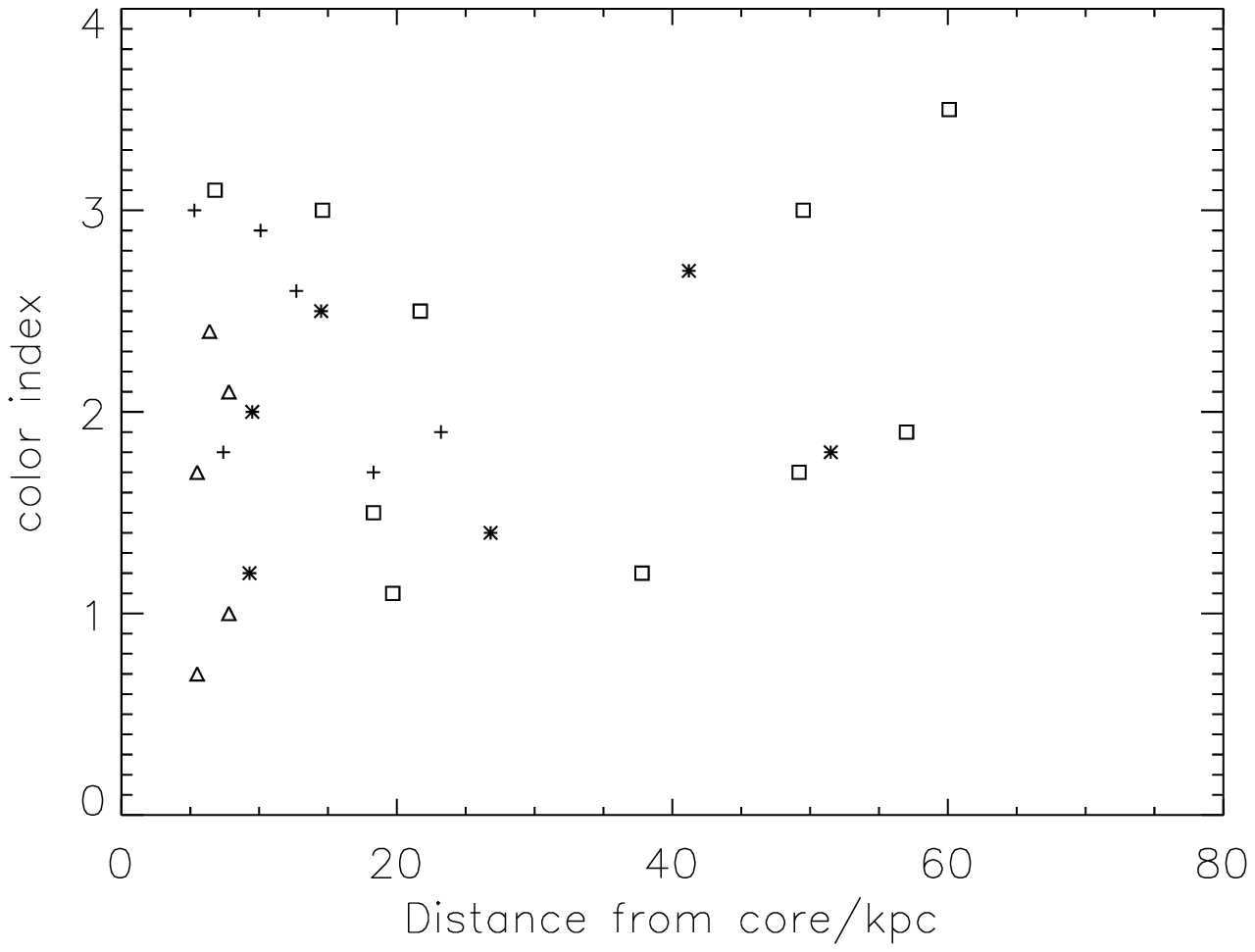}
\caption[f31.ps]{The optical-to-infrared color indices of the components
of the 4 clumpy radio galaxies MRC 0406$-$244 (triangles), MRC 1138$-$262 (squares), USS 1707$+$105 (asterisks) and MRC 2104
$-$242 (plus), plotted versus their distance from
the galaxy radio core (or nucleus).}\label{color}\end{figure}\begin{figure}
\plotone{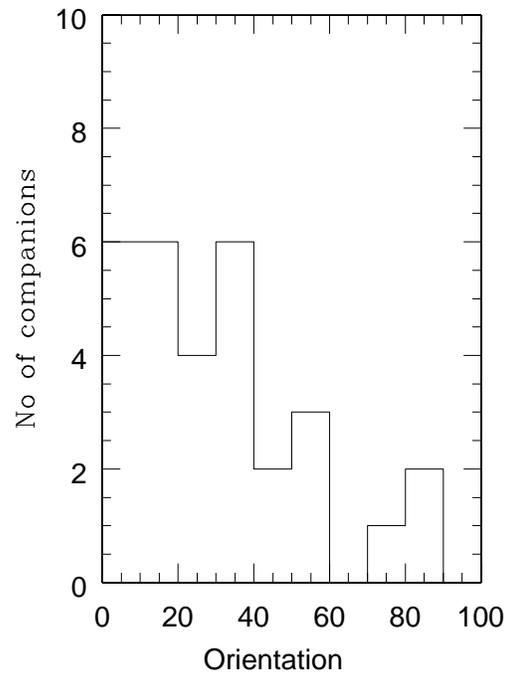}
\caption[f32.ps]{Distribution of the orientation of companion galaxies within 50 kpc of
the radio galaxies, with respect to the radio axis}\label{orient}
\end{figure}
\end{document}